\documentclass[a4paper,fleqn]{mnras}

\usepackage{epsfig,color}
\title[Recurring OH Flares towards $o$~Ceti]
       {Recurring OH Flares towards $o$~Ceti: I. location and structure of 
       the 1990s' and 2010s' events}

\author[Etoka {\em et al.}]
{S. Etoka$^{1}$\thanks{E-mail: Sandra.Etoka@googlemail.com},
E. G\'erard$^2$, A.M.S. Richards$^3$, D. Engels$^1$, 
        J. Brand$^4$, T. Le Bertre$^5$\\
$^1$ Hamburger Sternwarte, Gojenbergsweg 112, 21029 Hamburg, Germany \\
$^2$ GEPI, UMR 8111, CNRS \& Observatoire de Paris, 5 Place J. Janssen, F-92195 Meudon Cedex, France \\
$^3$ Jodrell Bank Centre for Astrophysics, School of Physics and Astronomy, University of Manchester, M13 9PL, UK \\
$^4$ INAF - Istituto di Radioastronomia \& Italian ALMA Regional Centre, Via P. Gobetti 101, 40129 Bologna, Italy \\
$^5$ LERMA, UMR 8112, CNRS \& Observatoire de Paris, 61 av. de l’Observatoire, F-75014 Paris, France \\
}

\date{Accepted 2017 February 20. Received 2017 February 15; in original form 2016 June 14.}

\pagerange{\pageref{firstpage}--\pageref{lastpage}} \pubyear{2017}

\begin{document}
\maketitle

%----------------------------------------------------------------------
\begin{abstract} 

We present the analysis of the onset of the new 2010s' OH flaring event 
detected in the OH ground-state main line at 1665~MHz towards $o$~Ceti and 
compare its characteristics with those of the 1990s' flaring event.
This is based on a series of complementary single-dish and interferometric  
observations both in OH and H$_2$O obtained with the {Nan\c cay} Radio 
telescope (NRT), the Medicina and Effelsberg Telescopes, 
the European VLBI Network (EVN), 
and (e)Multi-Element Radio Linked Interferometer Network ((\emph{e})MERLIN).
We compare the overall characteristics of $o$~Ceti's flaring events 
with those which have been observed towards other thin-shell Miras, 
and explore the implication of 
these events with respect to the standard OH circumstellar-envelope model. 
The role of binarity in the specific characteristics of $o$~Ceti's flaring 
events is also investigated. The flaring regions are found to be less than 
$\sim$400$\pm$40~mas (i.e., $\le 40 \pm 4$~AU) either side of $o$~Ceti,   
with seemingly no preferential location with respect to the direction to the 
companion Mira~B. 
Contrary to the usual expectation that the OH maser zone 
is located outside the H$_2$O maser zone, the coincidence of the H$_2$O and OH 
maser velocities suggests that both emissions arise at similar distances 
from the star. 
The OH flaring characteristics of Mira are similar
to those observed in various Mira variables before, supporting the
earlier results that the regions where the transient OH maser emission
occurs are different from the standard OH maser zone.
\end{abstract}

%- - - - - - - - - - - - - - - - - - - - - - - - - - - - - - - - - - - -

%----------------------------------------------------------------------
\section{Introduction}
\label{sec: Introduction}

Asymptotic Giant Branch (AGB) stars have a high mass-loss rate 
(typically 10$^{-7}$ -- 10$^{-5}$ M$_{\odot}$~yr$^{-1}$) leading to the 
creation of a dusty and molecular-rich circumstellar envelope (CSE).
SiO, H$_2$O, and OH masers are commonly emitted by the CSEs of 
oxygen-rich AGB Miras and OH/IR stars. These masers are a powerful tool to 
study the dynamical and structural evolution of the CSE while the star 
evolves towards the [proto-]planetary nebula stage, which in turn is crucial 
in understanding e.g., how asymmetries, commonly observed in the 
proto-planetary nebula stage but not so much during the AGB evolution, develop.

In standard models, the CSE is in spherical radial expansion with masers 
tracing an `onion-shell' structure (Omont 1988; Habing 1996). 
SiO is found closest to the star (typically within 4~R$_{*}$, where 
R$_{*} \sim$1~AU for a Mira), 
surrounded by H$_2$O (outside the radius where dust formation is complete) out 
to a few tens R$_{*}$, whilst OH, created by the photodissociation 
of H$_2$O by external ambient UV radiation, is found in the outer part of the 
CSE (typically $\ge$100R$_{*}$). 

Miras are pulsating stars with periods ranging between $\sim$100 and 
$\sim$500~days. 
Long-term monitoring observations towards Miras revealed that the standard OH 
maser emission, though exhibiting slow modulation which spreads over several 
cycles, varies smoothly lagging behind the optical curve by about 10-20\% of 
the period (Etoka \& Le~Squeren 2000). 
The polarisation of the overall emission, is typically $\le$~20\% for 
the 1665/67-MHz main lines and $\le$~10\% for the 1612-MHz satellite line 
(Wolak, Szymczak \& G\'erard, 2012). 

Towards optically thin-shell Miras with low mass-loss rate, 
OH masers have shown particularly unexpected behaviour in the form of flares, 
that is the sudden emergence and subsequent fading away of strong OH maser 
emission in one of the ground-state lines, which can persist for several years. 
These flares occur at velocities close to the stellar velocity, 
suggesting that they emanate from closer to the star compared to the distance 
at which standard OH emission arises in the CSE 
(Jewell et al. 1981, Etoka \& Le~Squeren 1996, 1997), but this has never 
been confirmed by imaging. \\

$o$~Ceti, Mira `the wonderful' has become synonymous with cool, pulsating
AGB stars. With a distance estimated by HIPPARCOS to be 92 $\pm$11~pc 
(and a proper motion of PM$_{\rm RA}= +9.33$~mas~yr$^{-1}$ and
PM$_{\rm Dec} = -237.36$ mas~yr$^{-1}$ van~Leeuwen 2007), it is one of the closest 
Miras. It has an estimated mass-loss rate in the range of  
$\sim$[1 - 4]~$\times$~10$^{-7}$~${\rm M_{\odot} yr^{-1}}$
(Knapp et al. 1998, Winters et al. 2003).
Because of its typical optical period of 332~days, which has been mo\-nitored 
for centuries, it is also known as the prototype of Mira long-period variable 
stars. It actually belongs to a detached binary system (Mira AB) in which 
mass transfer by wind interaction is taking place. 
Sokolski \& Bildsten (2010) found evidence pointing towards a white 
dwarf nature of Mira~B, while $o$~Ceti (Mira~A) shows clear signs of stellar 
asymmetry (Karovska et al. 1997; Reid \& Menten 2007).
$o$~Ceti and its companion Mira~B are separated by only $\sim$0.5$\arcsec$ 
(Karovska et al. 1997), corresponding to $\sim$46~AU. The orbital elements 
of the system, originally determined by Baize (1980) lead to an orbital 
period of P$\sim$400~yr. A more recent revision by Prieur et al. (2002) 
leads to an increase of the estimation of the period by nearly 100~yr 
(the new estimated period being P$\sim$498~yr).

$o$~Ceti is associated with persistent SiO and H$_2$O masers 
(Cotton et al. 2006; Bowers \& Johnston 1994). The first detection of OH 
maser emission in its CSE was made in 1974 
(Dickinson, Kollberg \& Yngvesson 1975) in the 1665-MHz ground-state main-line 
transition. The status of its OH ground-state maser lines at 1665, 1667 \& 
1612~MHz was checked with the NRT around an optical maximum phase in the 
late 1970's down to a sensitivity limit of 70~mJy by 
Sivagnanam, Le~Squeren \& Foy (1988) who reported it as non-emissive.

Redetection of its OH-maser emission in the 1665-MHz line was made in 
November 1990 (G\'erard \& Bourgois 1993). The emission was then monitored 
at the NRT until it faded away nearly a decade later, 
late 2000. The strongest emission recorded in 2000 
occurred on 23 November 2000 at phase $+0.21$ with a peak flux density of 
250~mJy at a radial velocity of 46.7~km/s, which is believed to be the 
trail of the 1990s flare. Between December 2000 and December 2008,  
scattered observations of $o$~Ceti were made between phases $-0.42$ and
$+0.45$ with 7 tentative detections at a level never exceeding 100~mJy. 
This does not exclude that the 1665-MHz level was ever present at a 
level $\leq$  50~mJy.
In November 2009 we detected with the NRT a new flare in the OH 1665-MHz line 
towards $o$~Ceti currently (2016) still active. \\

We present here a comparison of the 1990s' flaring event characteristics 
with those of the 2010s' event, based on a series of complementary 
single-dish and interferometric observations both in OH and H$_2$O with the 
{Nan\c cay}, Medicina and Effelsberg Radio Telescopes, 
MERLIN and EVN-(\emph{e})MERLIN.
The details of the observations are presented in
section~\ref{sec: observations}. The results of the OH and H$_2$O single-dish  
monitoring as well as the OH mappings obtained during the 1990s' event and 
around the first OH maximum recorded during the new 2010s' event 
are presented in section~\ref{sec: results}.
A discussion of the results is given
in section~\ref{sec: discussion}. Finally, a summary and conclusions are 
presented in section~\ref{sec: conclusion}.

%----------------------------------------------------------------------
\section{Observations and data reduction} 
\label{sec: observations}

\subsection{The NRT OH observations} 
\label{sub: NRT OH observations}

The NRT is a transit instrument with a half-power beamwidth of 
3.5$\hbox{$^\prime$ }$ in right ascension (RA) by 19$\hbox{$^\prime$ }$ in 
declination (DEC) at 1.6~GHz. \\

For the 1990s' observations, the system noise temperature was 45~K 
at ${\rm0^{\circ}}$ declination.
The antenna temperatures were converted to flux densities using
the efficiency curve of the radiotelescope which was 0.9~${\rm ~K~Jy^{-1}}$
for point sources at ${\rm0^{\circ}}$ declination. 
An autocorrelation spectrometer consisting of 4 banks of 256 channels
was used to observe both left-hand and right-hand circular (LHC and RHC) 
polarisations of the two ground-state OH main lines at 1665 and 1667~MHz, 
providing Stokes parameters I and V. 
A bandwidth of 0.0975~MHz was used for each bank leading to a velocity 
resolution of 0.0703~km~s$^{-1}$. 
Both linear vertical and horizontal polarisations (corresponding to the 
polarisation position angle PPA = $0^{\circ}$ and PPA = $90^{\circ}$, 
respectively) in the main lines were also regularly observed, providing the 
Stokes parameters I and Q. A typical observation, taken with a uneven sampling 
ranging from 1 day to $\sim$1.5~month, consisted in 40~minutes, taken in 
frequency switching mode, resulting in a mean rms of 100~mJy. \\

For the 2009/2010 observations presented here (corresponding to the time 
interval 2$^{d}$ December 2009~--~21$^{th}$ November 2010), 
the system temperature was about 35~K. The ratio of flux to antenna 
temperature was 1.4~${\rm ~K~Jy^{-1}}$ at ${\rm0^{\circ}}$ declination. 
The 8192 channel autocorrelator was divided into 8 banks of 1024 channels each. 
A bandwidth of 0.195~MHz was used for each bank, leading to a velocity 
resolution of 0.0343~km~s$^{-1}$. 
Prior and up to the first observation leading to the detection of the flare, 
observations were taken monthly over several years as part of a wider Key 
project. After the detection, the observations were taken every 5~days. 
Both ground-state main lines at 1665~MHz and 1667~MHz were recorded, and the 
ground-state satellite line at 1612~MHz was also regularly observed.
A series of 2 successive observations of 30~minutes each in frequency switching 
mode were taken so as to obtain full polarimetric observations for both OH 
ground-state main lines in a similar fashion as described in 
Szymczak \& G\'erard (2004) hence providing the 4 Stokes parameters 
(I, Q, U and V) and the 2 circular (LHC, RHC) polarisations. 
This integration time provided a mean rms of about 90~mJy 
(decreasing to 55~mJy by ``moving-average''
smoothing over 3 channels when the signal was faint, hence degrading slightly 
the velocity resolution). \\

The flux-density scale accuracy for both epochs is $\sim$10\%.

\subsection{The interferometric OH observations} 
\subsubsection{The MERLIN observations} 
\label{sub: MERLIN OH observations}

The MERLIN 1665-MHz OH phase-referenced interfe\-rometric observations 
obtained during the 1990s' flaring event were taken on the 28$^{th}$ of November 
1995 and the 4$^{th}$ of May 1998. 
For the 1995 observations, the 8 telescopes of MERLIN available at that time
(namely Defford, Cambridge, Knockin, Wardle, Darnhall, MK2, Lovell and Tabley)
were used. 
The source was observed with a bandwith of 0.125~MHz divided into 128 
channels at correlation, leading to a channel separation of 1~kHz 
(giving a velocity resolution of 0.18~km~s$^{-1}$). 
For the 1998 observations, only six telescopes were used: 
Defford, Cambridge, Knockin, Darnhall, MK2 and Tabley.
The source was observed with a bandwith of 0.25~MHz, divided into 256 channels 
at correlation, leading to the same channel separation as for the first epoch. 
For both epochs, J0219+0120, with a separation of 4.30$^\circ$ from the target, 
was used as the phase-reference calibrator,
3C286 was used as the primary flux density reference, and 
the continuum source 3C84 was used to derive corrections for instrumental 
gain variations across the band. 
For both epochs the same pointing position was used for the source: 
RA$_{\rm B1950}=02^h16^m49.062^s$ and 
DEC$_{\rm B1950}=-03^{\circ} 12\arcmin 22.480\arcsec$
(corresponding to 
RA$_{\rm J2000}=02^h19^m20.771^s$ and 
DEC$_{\rm J2000}=-02^{\circ} 58\arcmin 36.674\arcsec$). \\

The initial editing, the gain-elevation correction and a first-order 
amplitude calibration of the MERLIN datasets were applied using the 
MERLIN-specific package {\tt DPROGRAMS} while the second order 
calibration and further data processing analysis were performed with the 
{\tt AIPS} package, following the procedure explained in section~2.2 
of Etoka \& Diamond (2004).
The flux-density scale accuracy is estimated to be $\sim$5\%.

\subsubsection{The EVN-(\emph{e})MERLIN observations} 
\label{sub: EVN-(e)MERLIN OH observations}

The EVN-(\emph{e})MERLIN OH phase-referenced in\-ter\-fe\-ro\-me\-tric 
observations were obtained just past the first OH maximum after the onset of 
the new 2010s' flaring event, the 10$^{th}$ of February 2010. 
Eight telescopes were used, namely: 
Effelsberg,  Lovell, Westerbork, Onsala, Medicina, Noto, Toru\'n and Cambridge. 
These observations were taken in full polarisation
spectral mode though only the relative PPAs could be 
determined as the observations of the PPA calibrator 3C286 were 
unfortunately unsuccessful, ruling out the determination of the absolute PPA. 
The source was observed with a bandwidth of 2~MHz divided into 2048 channels 
at correlation leading to a channel separation of 1~kHz 
(giving a velocity resolution of 0.18~km~s$^{-1}$). The pointing position 
used for the source was: 
RA$_{\rm J2000}=02^h19^m20.790^s$ and 
DEC$_{\rm J2000}=-02^{\circ} 58\arcmin 41.830\arcsec$.
The continuum source J0237+2848 was used as a fringe finder.
The continuum source 3C84 was used to derive corrections for instrumental 
gain variations across the bandpass and correct for polarisation leakage. 
J0217-0121 which was used as the phase-reference calibrator, is located 
1.65$^\circ$ away from the target. \\

The EVN-(\emph{e})MERLIN dataset editing and calibration and subsequent 
imaging were performed entirely with {\tt AIPS}. 
The EVN pipeline was used for standard initial steps including deriving 
amplitude calibration from the system temperature monitoring and the 
parallactic angle correction. We then followed usual VLBI procedures for 
spectral line observations including solving for delay residuals and the
time-derivative of phase, followed by bandpass calibration and the application 
of phase reference solutions to the target.

The flux-density scale accuracy is estimated to be $\sim$10\%.

\subsubsection{Accuracy of the absolute and relative positions} 
\label{sub: position accuracy}

Regarding the absolute positional accuracy of the interferometric observations, 
adding up quadratically all the factors affecting the positional accuracy 
(i.e., the positional accuracy of the phase-reference calibrator, the 
accuracy of the telescope positions, the relative positional error depending on 
the beamsize and signal-to-noise ratio (SNR) and finally the atmospheric 
variability due to the angular separation between the phase-reference 
calibrator and the target) leads to an estimated total uncertainty of the 
absolute position of (35$\times$25)~mas$^2$ for the MERLIN observations and 
(35$\times$10)~mas$^2$ for the EVN-(\emph{e})MERLIN observations, in RA and 
DEC respectively.  
The absolute positions were measured before self-calibrations.
The relative positional accuracy itself, given approximately by the 
beamsize/SNR (Thompson, Moran \& Swenson 1991; Condon 1997; 
Richards, Yates \& Cohen 1999) is typically down to a few mas. 
It has to be noted that, when comparing observations made with the same array
and phase reference source, the relative positions are only affected
by the noise and atmospheric errors. \\

\subsection{The Medicina and Effelsberg H$_2$O observations} 
\label{sub: Medicina H2O observations}

$o$~Ceti was observed with the Medicina\footnote{The 32-m Medicina telescope 
is operated by the INAF-Istituto di Radioastronomia (IRA) at Bologna, Italy.} 
32-m antenna, as part of a larger programme, 3 to 5 times a year between 
December 1995 and March 2011 (68 spectra). 
In addition, 6 spectra were taken with the 
100-m dish at Effelsberg\footnote{The 100-m Effelsberg telescope is operated 
by the Max-Planck-Institut f\"ur Radioastronomie (MPIfR) at Bonn, Germany.} 
between March 1995 and February 1999.
All spectra will be presented by Brand et al. (2017).

At Medicina (beam FWHM$\approx$1.9\arcmin\ at 22~GHz), observations were made 
in total power mode with both ON and OFF scans of 5-min duration. 
The OFF position was taken 1$^\circ$ West of the source position. Typically, 
2 ON/OFF pairs were taken. The backend was a 1024-channel autocorrelator 
with a bandwidth of 10~MHz. The typical rms noise-level in the final spectra 
is $\sim 1.4$~Jy, for a velocity resolution of 0.13~km~s$^{-1}$ 
($\Delta \nu = 9.77$~kHz). For line observations at 22~GHz, only the LHC 
polarisation output from the receiver was registered. 

At Effelsberg (beam FWHM$\approx$40\arcsec\ at 22~GHz) the receiver passed 
only LHC polarised radiation. The backend consisted of a 
1024-channel autocorrelator with a bandwidth of 6.25~MHz, resulting in a 
velocity resolution of 0.08~km~s$^{-1}$. We observed in total power mode, 
integrating ON and OFF the source, for 5 min each. The ``OFF-source'' position 
was displaced 3\arcmin\ to the east of the source. The typical rms 
noise-level in the spectra is $\sim 0.2$~Jy. \\

All the radial velocities given hereafter are relative to the Local Standard 
of Rest (LSR). 

%----------------------------------------------------------------------
\section{Results}
\label{sec: results}

%-- -- -- -- -- -- -- -- -- -- -- -- -- -- -- -- -- -- -- -- -- -- -- -- -- --
\subsection{The flaring event of the 1990s}
\subsubsection{The single-dish NRT OH observations}
\label{sub: 1990 NRT obs}
\noindent
{\it 1665~MHz} \\
\noindent
The OH flare of the 1990s was monitored with the NRT over its
approximately ten-year duration. This paper focuses on comparison with
the MERLIN imaging of the flaring region 
(cf. section~\ref{sub: 1990 MERLIN maps}) so, here, we present only 
the spectra taken for a 30-days period around the MERLIN observations.
A detailed comparison of the spectral and variability characteristics of the 
overall period with the 2010s' flaring period will be presented in a 
subsequent paper.  \\

\begin{figure}
%\hspace*{-0.15cm} 
 \epsfig{file=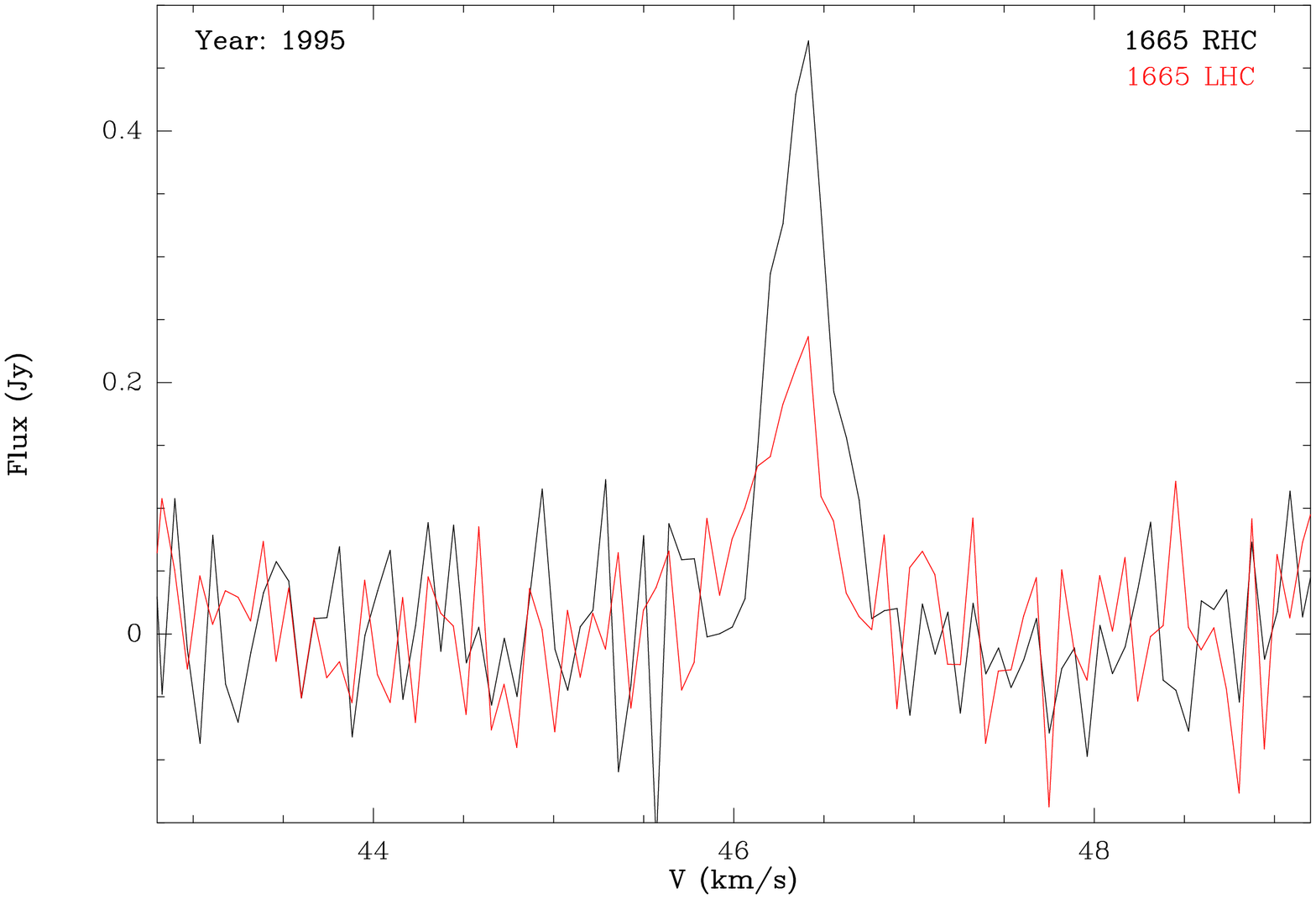,width=5.70cm,angle=-90}
\caption{Average of the 4 OH 1665-MHz single-dish observations obtained with 
 the NRT (over a period of 30 days, between mid-November and mid-December) 
 around the MERLIN observations of the 28$^{th}$ of November 1995.}
\label{fig: NRT 1665 in 95}
\end{figure}

\begin{figure}
\hspace*{-0.05cm} 
 \epsfig{file=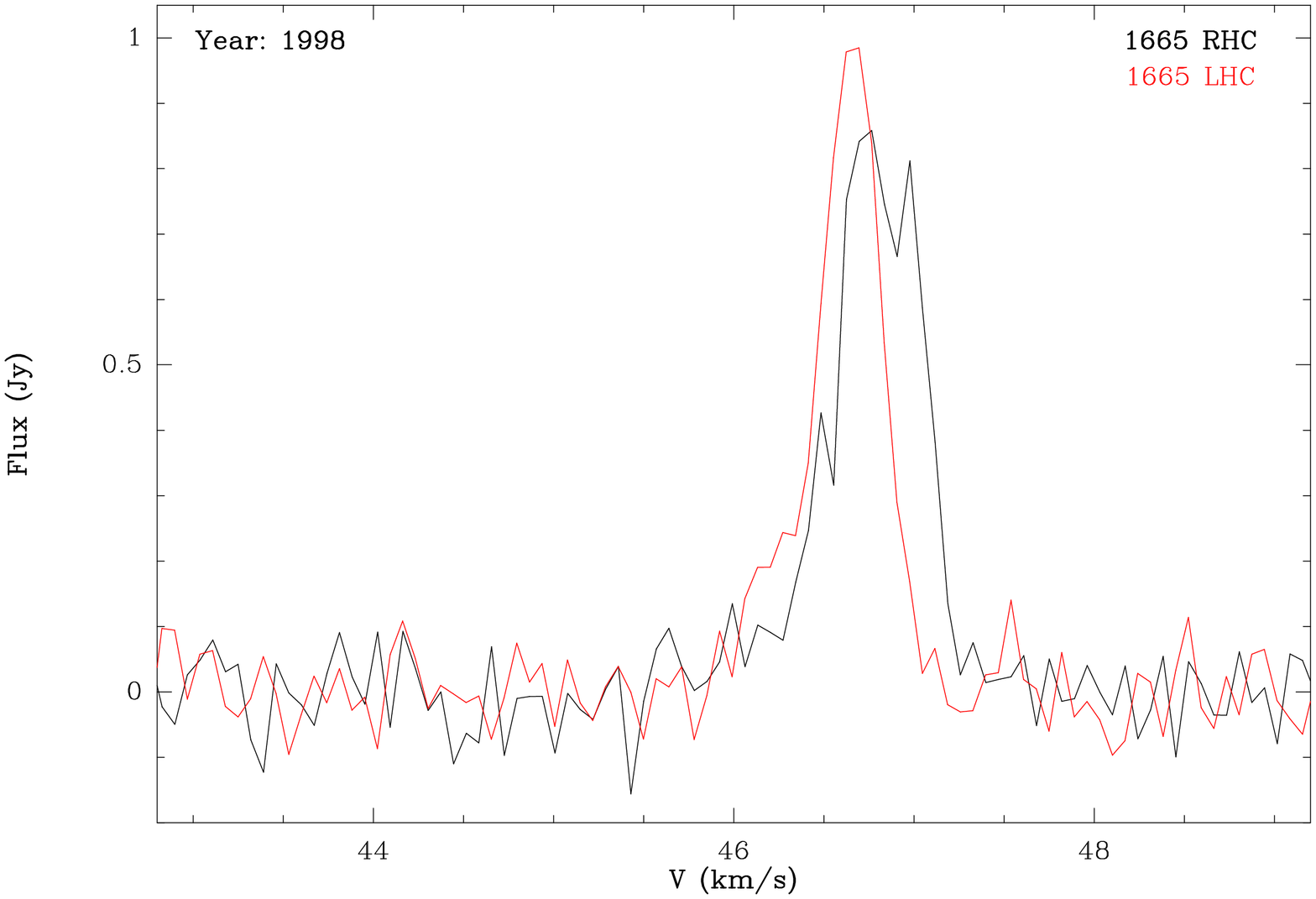,width=5.70cm,angle=-90}
\caption{Average of the 5 OH 1665-MHz single-dish observations obtained with 
 the NRT (over a period of 30 days, between mid-April and mid-May) around the 
 MERLIN observations of the 4$^{th}$ of May 1998.}
\label{fig: NRT 1665 in 98}
\end{figure}

Figures~\ref{fig: NRT 1665 in 95}~\&~\ref{fig: NRT 1665 in 98} present 
2 single-dish spectra obtained with the NRT for the period closest to 
the MERLIN observations obtained on the 28$^{th}$ of November 1995 and 
the 4$^{th}$ of May 1998. Note that the NRT spectra presented here 
are an average of several observations taken around the 2 previously 
mentioned dates over a period of 30 days (between mid-November and 
mid-December 1995 and between mid-April and mid-May 1998 for the 
2 respective epochs) so as to increase the SNR.
The observations of the 28$^{th}$ of November 1995 (coinciding with an 
optical phase of $-0.32$) correspond to the beginning of an OH cycle 
(characterised by a gentle rise in intensity, showing moderate intrinsic 
variation) which roughly culminated early 
April 1996, while the observations taken on the 4$^{th}$ of May 1998 
(coinciding an optical phase of $+0.37$) correspond to the steep 
decreasing part towards an OH minimum, the maximum having been reached between 
mid-February and mid-March 1998. Therefore, it has to be noted that the peak 
intensity observed in the 1998 averaged-spectrum is dominated by the 
mid-April spectra which have the highest SNR.
The intensity of the emission inferred from the NRT 
spectra obtained the clo\-sest to the MERLIN observations 
(i.e, mid-May spectra) attests of F$_{peak \; LHC} \sim 0.73$~Jy and  
F$_{peak \; RHC} \sim 0.65$~Jy.

Though the overall profile characteristics in terms of velocity 
spread and the presence of at least 2 components in the profile are similar 
between the 2 epochs, polarimetric changes in the region are clearly present 
with an inversion in the strength of the LHC and RHC. Also noticeable is a 
drift in the peak velocity between these 2 periods with 
V$_{\rm peak _{\, LHC \; 1995}} \simeq$~V$_{\rm peak _{\, RHC \; 1995}} \sim +46.42$~km~s$^{-1}$ 
while V$_{\rm peak _{\, LHC \; 1998}} \sim +46.69$~km~s$^{-1}$  and
V$_{\rm peak _{\, RHC \; 1998}} \sim +46.77$~km~s$^{-1}$. This corresponds to a 
velocity drift of about $[+0.27,+0.35]$~km~s$^{-1}$ in 888 days. \\

%%%%%%%%%%%%
%%%%%%%%%%%%
%%%%%%%%%%%%
%%%%%%%%%%%%

%%%%%%%%%%%%
%%%%%%%%%%%%
%%%%%%%%%%%%
%%%%%%%%%%%%

\begin{figure}
\hspace*{-0.05cm} 
 \epsfig{file=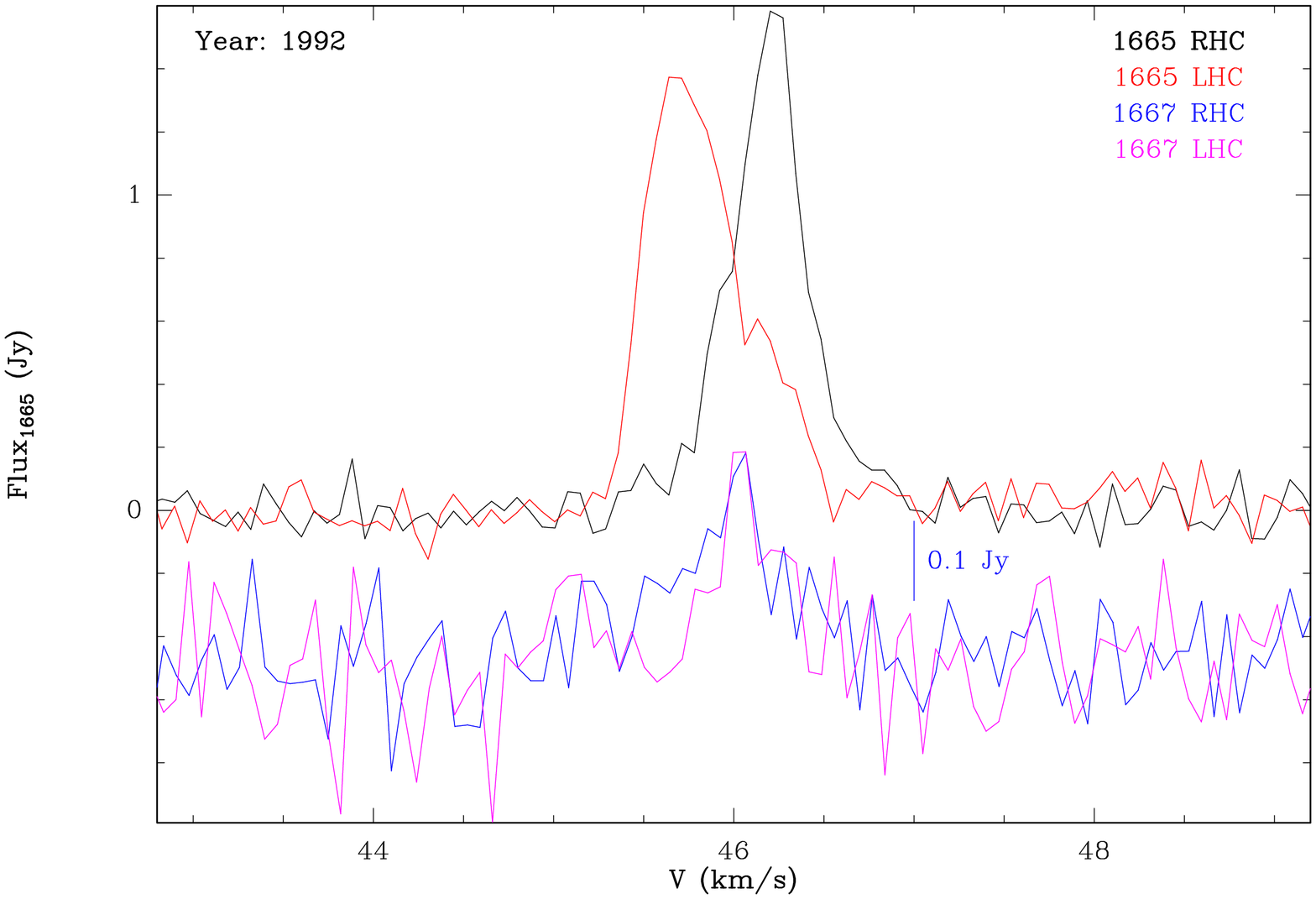,width=5.70cm,angle=-90}
\caption{Average of the 3 OH 1665-MHz \& 1667-MHz single-dish observations 
         obtained with the NRT between the 1$^{st}$ and the 10$^{th}$ of 
         October 1992 so as to increase the signal to noise.
         Note the difference in flux scale between the 1665~MHz spectra 
         (given by the left-hand-side coordinate axis) and the 1667~MHz spectra 
         (given by the vertical bar on the right-hand side of the LHC \& RHC 
         spectra themselves). }
\label{fig: NRT 1667 in 92}
\end{figure}

\noindent
{\it 1667~MHz} \\
\noindent
Faint 1667~MHz emission was detected intermittently. 
Figure~\ref{fig: NRT 1667 in 92} presents the detection made of this 
line, around the OH maximum of 1992 (at the optical phase +0.2) 
along with the corresponding 1665~MHz spectra for comparison. 
The figure is an average of the 3 NRT observations taken between the 
1$^{st}$ and the 10$^{th}$ of October 1992 so as to increase the SNR 
(note that the number of spectra selected for the average 
for the 3 series of spectra presented in Figs~\ref{fig: NRT 1665 in 95}, 
\ref{fig: NRT 1665 in 98}~\&~\ref{fig: NRT 1667 in 92} was chosen so as to get 
a similar rms of 55-60mJy, bringing out of the noise rather faint but 
reasonably long-lasting components). The LHC and RHC 1667-MHz peaks have 
similar intensities and are both centred at 
V$_{\rm peak} \sim +46.07$~km~s$^{-1}$, which is roughly the mid-point between 
the 1665-MHz LHC and RHC spectra in 1992 and is 
$\sim$0.35~km~s$^{-1}$ ``bluer'' than the 1665~MHz peak of the late-1995 cycle.
With a separation of 965~days, this hints towards a similar velocity drift 
over a similar time interval as the one recorded between late-1995 and 
mid-1998. \\

%%%%%%%%%%%%
%%%%%%%%%%%%
%%%%%%%%%%%%
%%%%%%%%%%%%

%- - - - - - - - - - - - - - - - - - - - - - - - - - - - - - - - - - - - - - -
\subsubsection{The MERLIN maps}
\label{sub: 1990 MERLIN maps}
\noindent

%%%%%%%%%%%%
%%%%%%%%%%%%
\begin{figure}
  \epsfig{file=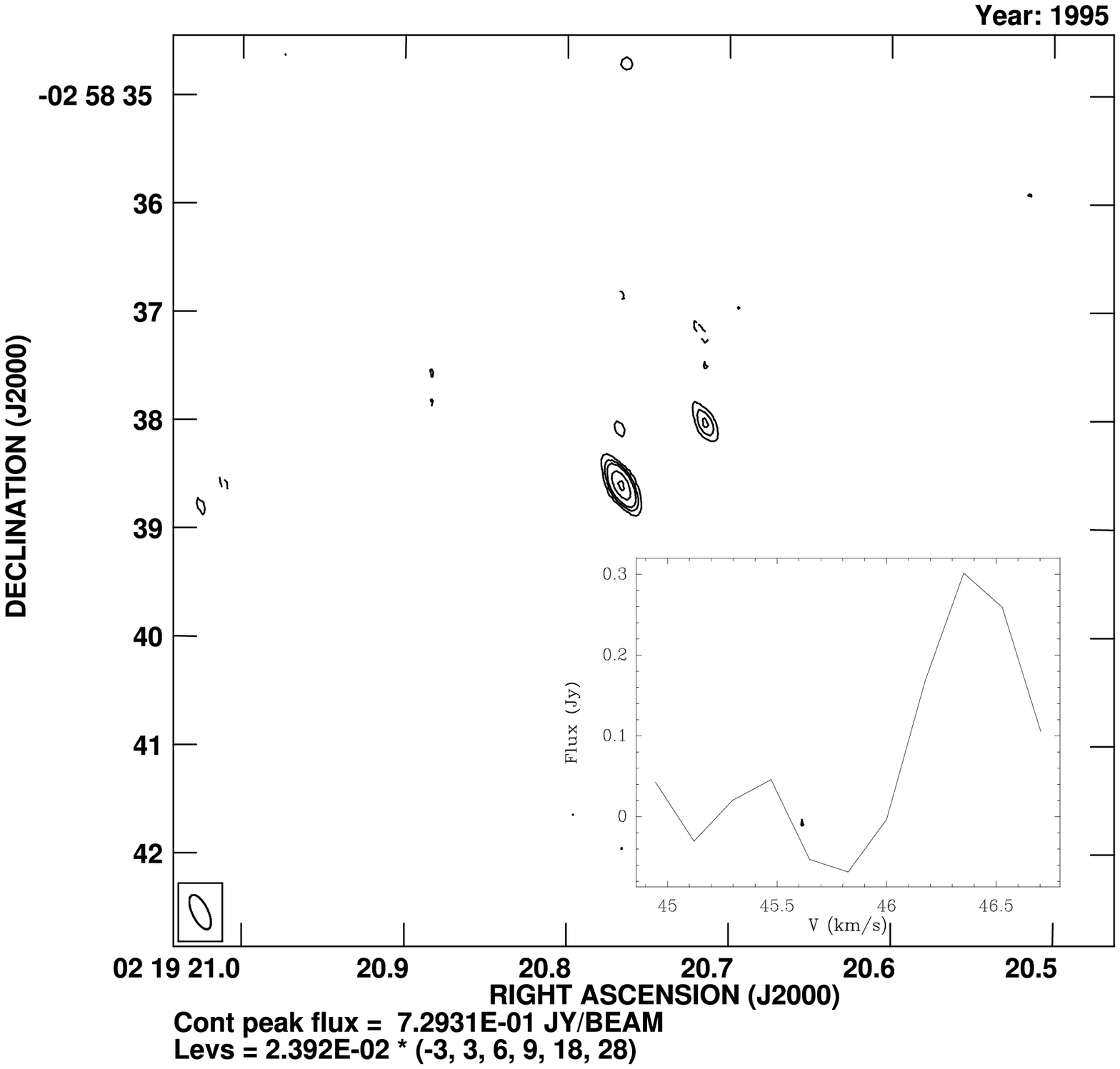,width=7.90cm,angle=-90}
\caption{1665-MHz OH maser map of the sum of all the channels where 
         emission has been detected in the Stokes~$I$ datacube 
         from the MERLIN observations taken on the 28$^{th}$ of November 1995. 
         The restoring beam (0.34610~$\times$~0.14505~arcsec$^2$, PA=$+26.22$) 
         is shown in the lower-left corner. The spectrum constructed from the 
         Stokes~$I$ datacube is shown in superimposition in the bottom-right 
         corner of the map.}
\label{fig: MERLIN 95}
\end{figure}
%%%%%%%%%%%%
%%%%%%%%%%%%

%%%%%%%%%%%%
%%%%%%%%%%%%
\begin{figure}
  \epsfig{file=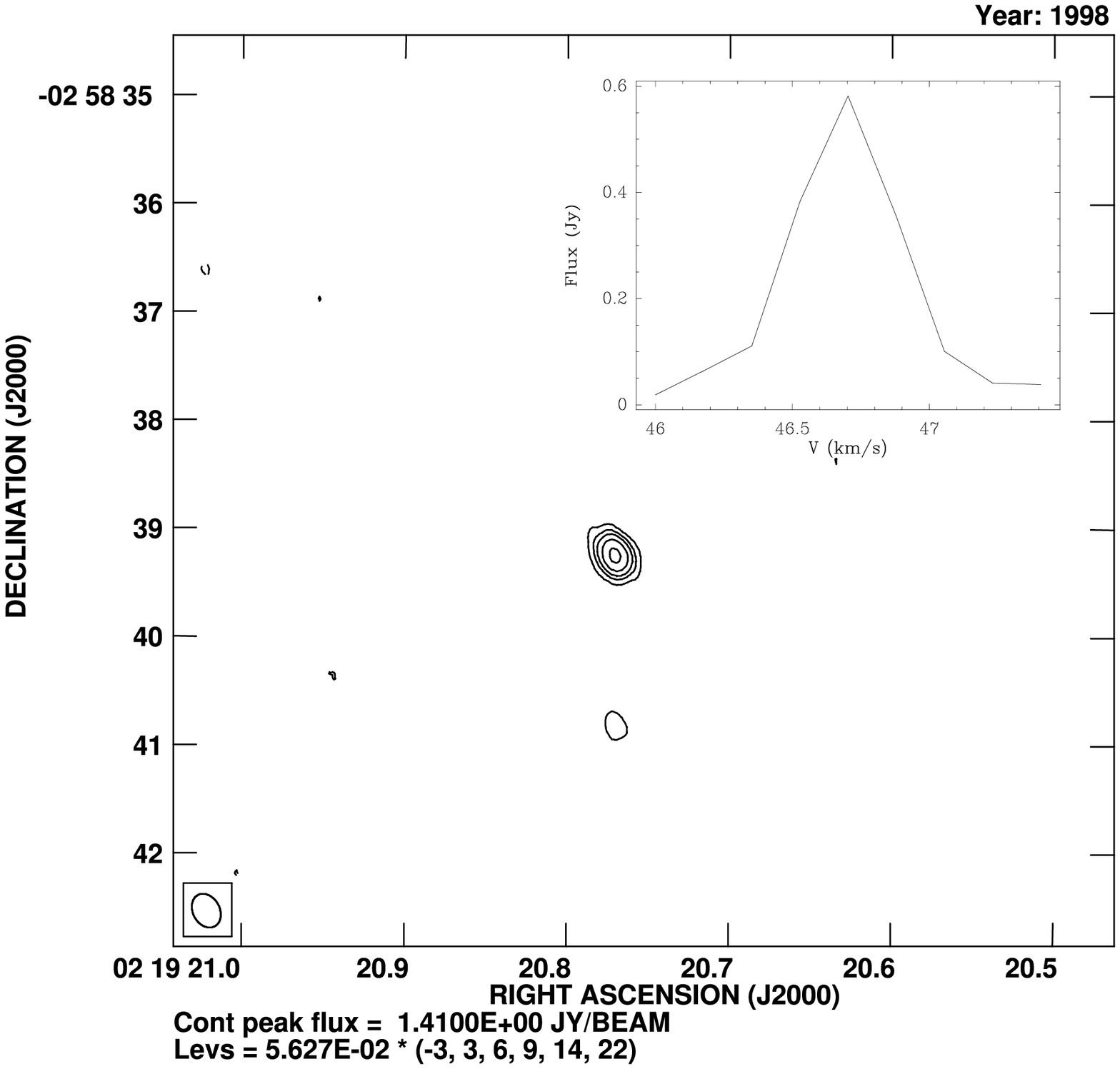,width=7.90cm,angle=-90}
\caption{Same as for Fig.~\protect\ref{fig: MERLIN 95} but for the MERLIN 
         1665-MHz OH maser observations taken on the 4$^{th}$ of May 1998. 
         The restoring beam (0.32781~$\times$~0.24763~arcsec$^2$, PA=$+28.66$) 
         is shown in the lower-left corner while the spectrum constructed from 
         the Stokes~$I$ datacube is superimposed in the top-right corner of 
         the map.}
\label{fig: MERLIN 98}
\end{figure}
%%%%%%%%%%%%
%%%%%%%%%%%%

Two epochs of interferometric observations with MERLIN were taken in 1995 and 
1998, 2 years and 5 months apart.
For these 2 datasets, the full polarisation information was not retrievable. 
Since no (circularly-polarised) substructure can be seen in the maser 
components themselves, only  
the resulting Stokes~$I$ maps are presented here in 
Figs~\ref{fig: MERLIN 95}~\&~\ref{fig: MERLIN 98}. These maps have been 
created by summing up all the channels where emission has been detected in the
final Stokes~$I$ datacube. The spectra built from these datacubes are also 
presented. The comparison of the peak intensity of the MERLIN spectra with the
NRT single-dish spectra (cf. Fig.~\ref{fig: NRT 1665 in 95}) implies that 
no significant part of the signal was missed during the interferometric 1995
observations. 

With the warning given in Section~\ref{sub: 1990 NRT obs} regarding the peak 
intensities of the averaged spectrum presented in 
Fig.~\ref{fig: NRT 1665 in 98} being biased towards the mid-April spectra, 
the comparison of the Stokes~$I$ peak intensity of the 1998 MERLIN spectra 
with that recorded mid-May by the NRT (i.e., the closest NRT spectra to the 
MERLIN observations which equate to F$_{peak \; (Stokes \;I)} \sim 0.69$~Jy) 
indicates that at least 85\% of the signal was recovered during the 1998 
interferometric observations.\\

In 1995, two maser components were detected se\-parated by about $1\arcsec$
(at RA$_{\rm J2000}$=$02^h19^m20.7647^s$ and 
    DEC$_{\rm J2000}$=$-02^{\circ} 58\arcmin 38.536\arcsec$ 
for the strongest, and
     RA$_{\rm J2000}$=$02^h19^m20.7131^s$ and 
     DEC$_{\rm J2000}$=$-02^{\circ} 58\arcmin 37.956\arcsec$ for the faintest), 
while only one maser component was detected in 1998 
(at RA$_{\rm J2000}$=$02^h19^m20.7687^s$ and 
    DEC$_{\rm J2000}$=$-02^{\circ} 58\arcmin 39.182\arcsec$).
The comparison of the proper-motion corrected positions of the 2 maser 
components detected in 1995 with the 1998 one is such that the strongest 
1995 component is located the closest to the 1998 component. 
Between the 2 epochs of observations, the difference in position of the 
strongest maser component 
has been measured to be $\delta_{\rm RA} \simeq +58.5$~mas  
$\delta_{\rm Dec} \simeq -680$~mas, while a proper motion of 
$\delta{\rm PM}_{\rm RA} \simeq +22.5$~mas and 
$\delta{\rm PM}_{\rm Dec}\simeq -574$~mas 
is expected. 
This leads to an actual positional difference of $\simeq$110~mas 
($\sim$13~AU) between them.
Following the V$_{\rm exp}$=f(period) relation for Miras of 
Sivagnanam et al. (1989), and taking into account the spread in this relation, 
a standard OH expansion velocity of 3-5~km\,s$^{-1}$ is expected for $o$~Ceti. 
The same maser component would then be expected to have travelled a linear 
distance of $\sim$1.5--2.5~AU radially outward over the 888~days separating 
the 2 MERLIN observations. 
This expected propagation is smaller by at least a factor of 5 than the 
measured separation between the 2 strongest maser components. Also, the 
measured separation is greater than the absolute positional uncertainty which 
means that if those 2 maser components are probing the same region of the 
shell, as expected since both epochs correspond to the same event evolving 
over a time interval of $\sim$10~yr, then the difference in position observed 
here represents a genuine snapshot of the propagation of the flare within the 
affected region, the disturbance speed being $\sim$25~km\,s$^{-1}$. \\

%-- -- -- -- -- -- -- -- -- -- -- -- -- -- -- -- -- -- -- -- -- -- -- -- -- -- 

\subsection{The flaring event of the 2010s}
\subsubsection{The single-dish NRT OH monitoring}
We present here the status of the observations obtained during the 
2009-2010 cycle, which corresponds to the first cycle of the 2010s' 
OH flaring event. All 3 ground-state lines known to be potentially present 
towards O-rich evolved stars, that is the 2 main lines at 1665 and 1667~MHz and 
the satellite line at 1612~MHz were monitored. No emission was detected in 
either the 1667- or the 1612-MHz lines. \\

\begin{figure}
\epsfig{file=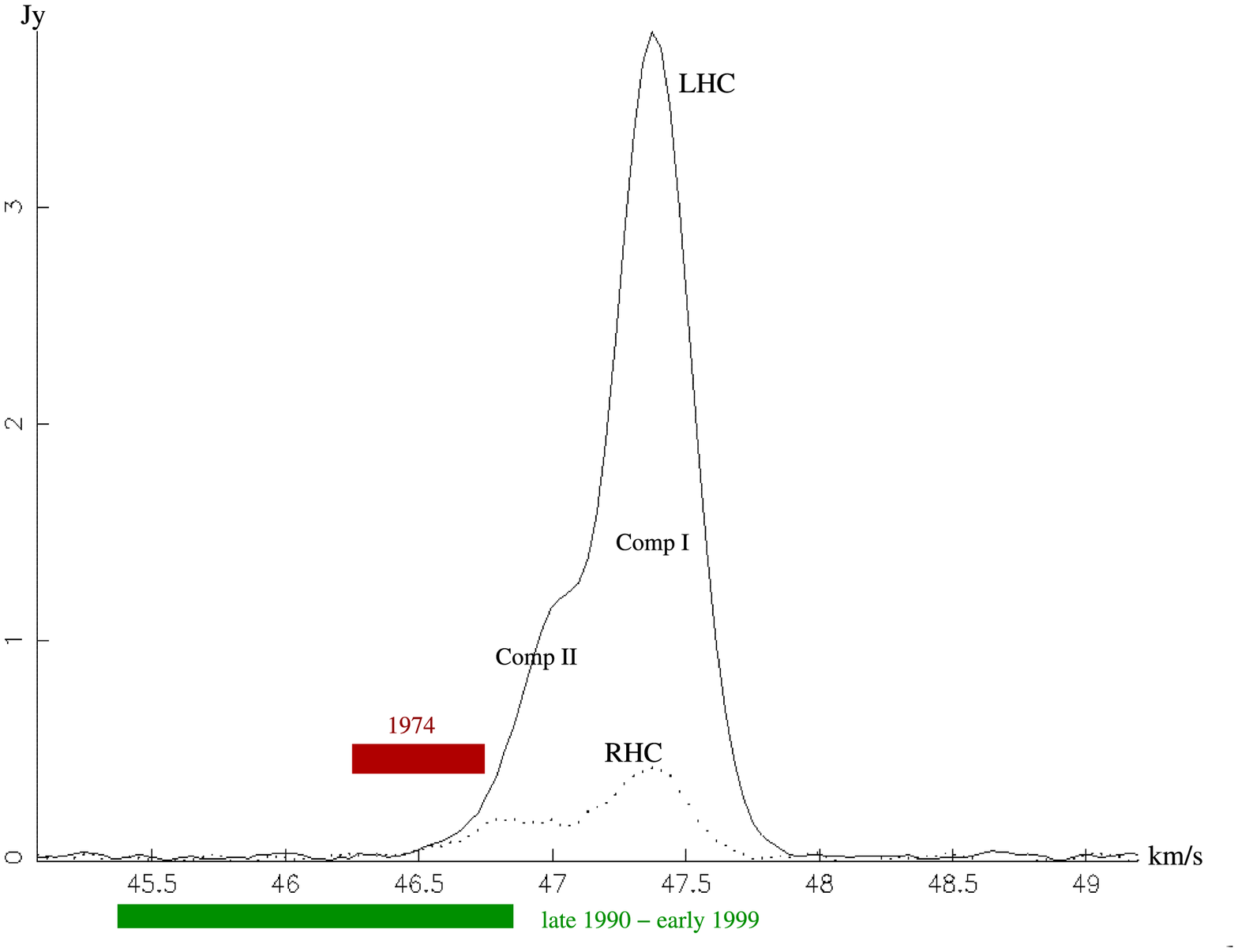,width=6.90cm,angle=-90} 
\caption{Average NRT spectra  in the LHC (solid line) and RHC (dotted line) 
         polarisations covering the period December 2009~--~February 2010, 
         (corresponding roughly to the MJD interval [55170--55260]) which 
         encompasses the OH maximum of the first cycle of the 2010s' OH flaring 
         event. 
         The red horizontal bar gives the velocity range of the OH 
         emission recorded in 1974, and the green horizontal bar gives the 
         OH emission velocity spread in 1990-1999. The 2 persistent spectral 
         components observed are labelled Comp~I and Comp~II for the 
         strongest and the faintest one respectively.}
\label{fig: NRT Inte OH 1665}
\end{figure}

Figure~\ref{fig: NRT Inte OH 1665} shows the average NRT spectrum of the
first cycle of the 2010s' OH flaring event observed in the 1665-MHz line.
The average spectrum is made of all 
the spectra taken between December 2009 and February 2010 inclusive 
(corresponding roughly to the modified Julian day (MJD) interval 
[55170--55260]), that is the spectra with the 
highest SNR since encompassing the OH maximum. 
The velocity information of the first detection of OH emission in 1974 
by Dickinson, Kollberg \& Yngvesson (1975) and the overall velocity span 
observed during the 1990s' flaring event (G\'erard \& Bourgois 1993)
are represented by the red and green horizontal bars, respectively. The new 
emission is composed of 2 strongly polarised main spectral components 
(the strongest and faintest components are labelled Comp~I and Comp~II, 
respectively). It presents a profile and a 
velocity spread similar to what was observed in the 1990s' flare but it is now 
centred at V$\simeq +47.1$~km~s$^{-1}$ (and spans the velocity interval 
V=[$+46.4$,$+47.8$]~km~s$^{-1}$), while in the 1990s, the flaring emission was 
centred at V$\simeq +46$~km~s$^{-1}$ (and spanned the velocity interval 
V=[$+45.3$,$+46.8$]~km~s$^{-1}$, cf. {G\'erard} \& Bourgois 1993, their Fig.~1). 
Note that Dickinson, Kollberg \& Yngvesson (1975) report the following 
characteristics for their observation in the 1970s: 
V$_{peak}$=$+46.5$~km~s$^{-1}$ with a line width $\Delta$V=0.5~km~s$^{-1}$.

With only one peak observed in the OH spectrum, it is likely that only one 
side of the shell is experiencing the flare as standard Miras commonly exhibit 
a double-peak profile with a typical expansion velocity of few ($<$10) 
km~s$^{-1}$ (Sivagnanam et al. 1989). \\

%%%%%%%%%%%%
%%%%%%%%%%%%
%%%%%%%%%%%%
%%%%%%%%%%%%

\begin{figure}
\epsfig{file=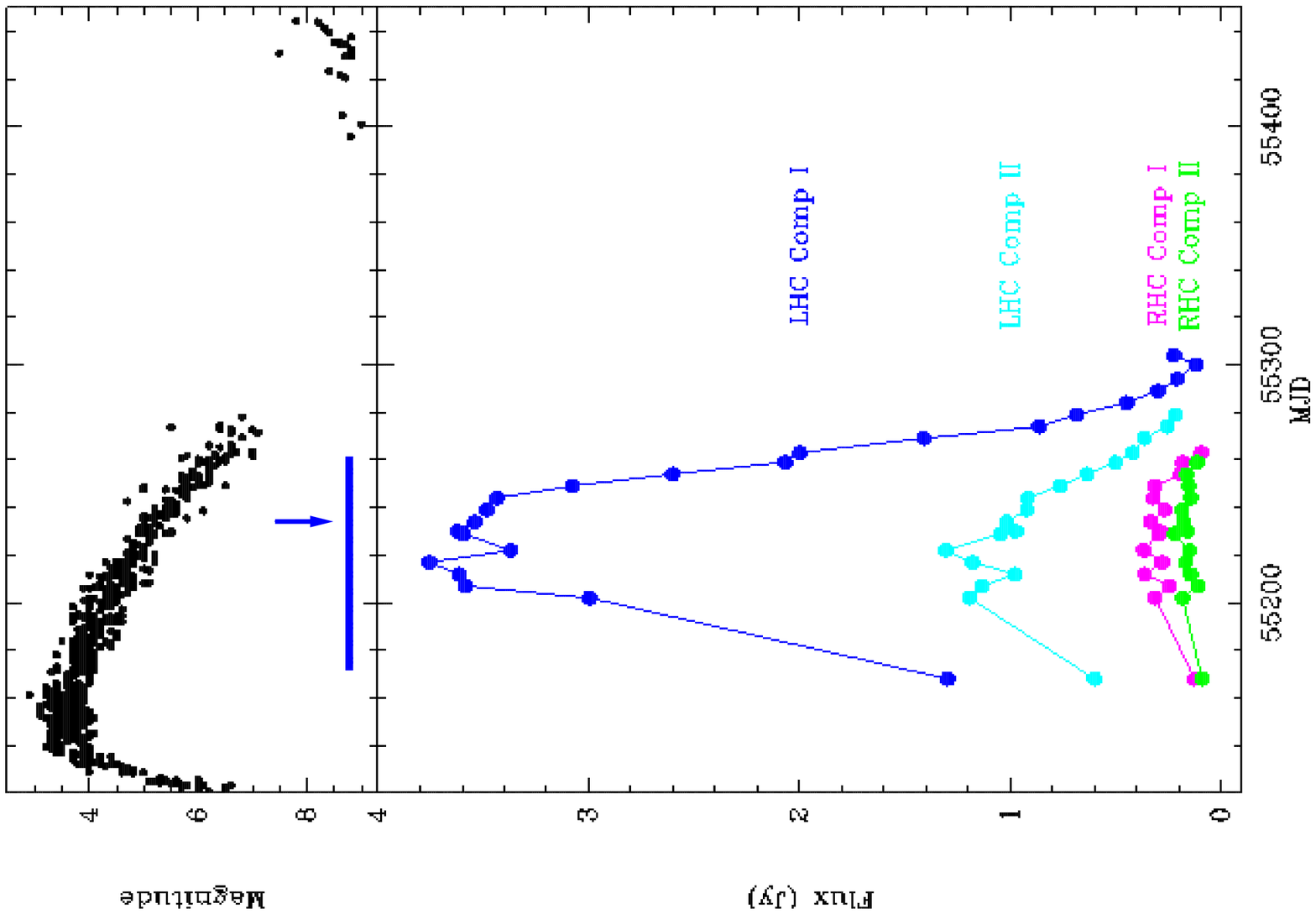,angle=-90,width=8.60cm}
\caption{{\bf Top panel:} $o$~Ceti's optical lightcurve of the 
         2009~--~2010 cycle (data from AAVSO).
         {\bf Lower panel:} Variability curves of the 1665-MHz individual 
         components (Comp~I and Comp~II as identified in 
         Fig.~\protect\ref{fig: NRT Inte OH 1665}) after Gaussian spectral 
         decomposition both in RHC and LHC polarisations. 
         The horizontal bar shows the interval over which the NRT spectrum 
         presented in Fig.~\protect\ref{fig: NRT Inte OH 1665} is averaged. 
         The arrow indicates the date corresponding to the EVN-(\emph{e})MERLIN 
         observations (taken at the optical phase +0.26). Note that 
         from MJD=55310 the signal decreased below 
         the noise level for the rest of the cycle presented here.}
\label{fig: NRT OH 1665 all component}
\end{figure}

Figure~\ref{fig: NRT OH 1665 all component} presents the variability curves 
of the indivi\-dual components after Gaussian spectral decomposition both in 
RHC and LHC polarisations. Note that from MJD=55310, the signal decreased 
below the noise level for the rest of the cycle presented here. These main 
components all follow the cycle with the expected delay with respect to the 
optical light curve, which is of $\sim$70 days corresponding to an optical 
phase of +0.2. 

%%%%%%%%%%%%
%%%%%%%%%%%%
%%%%%%%%%%%%
%%%%%%%%%%%%

%- - - - - - - - - - - - - - - - - - - - - - - - - - - - - - - - - - - - - - -
\subsubsection{The EVN-(\emph{e})MERLIN maps}
\label{sub: EVN-eMERLIN maps}

\begin{figure}
\epsfig{file=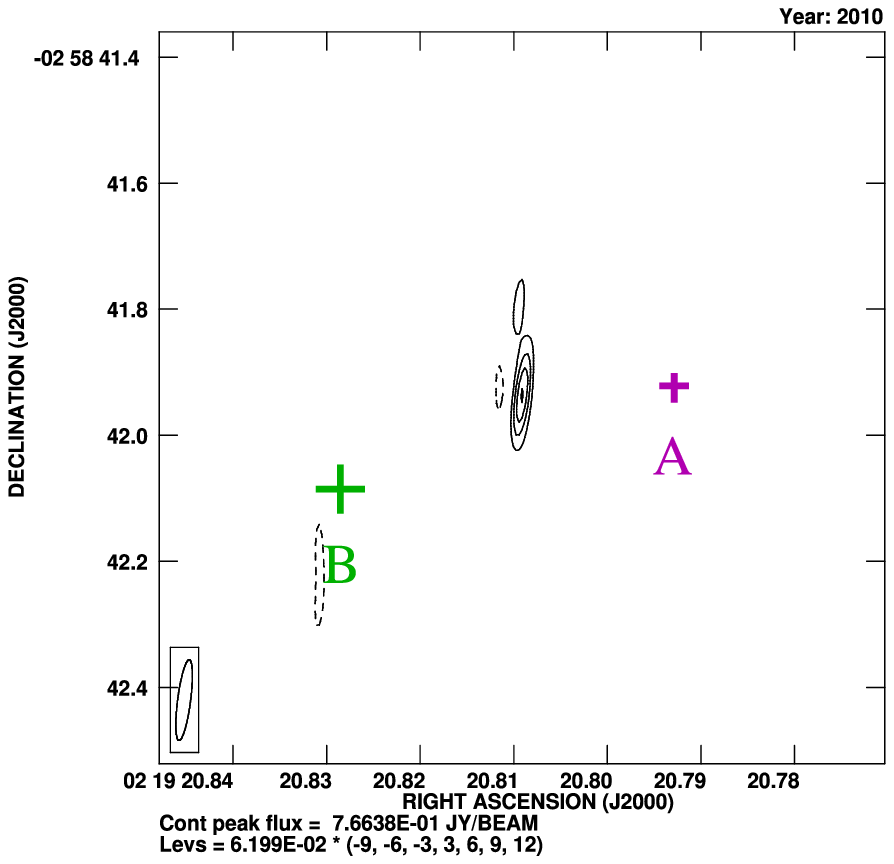,width=8.10cm,angle=-90}
\caption{Phase-referenced pre-self-calibration image of the 
  EVN-(\emph{e})MERLIN LHC map of the flare taken in February 2010 at the 
  optical phase +0.26. The restoring beam 
  (0.11770~$\times$~0.02044~arcsec$^2$, PA=$-13.62$) is shown in the 
  lower-left.
  The purple and green crosses give the estimated positions of $o$~Ceti 
  and its companion Mira~B, respectively. The positions of Mira~A is  
  extrapolated from Matthews \& Karovska's (2006) VLA imaging, 
  taking into account the proper motion (van~Leeuwen 2007). The 
  position of Mira~B is relative to that of Mira~A and was inferred from 
  the measurement values compiled by Planesas, Alcolea \& Bachiller (2016).
  The size of the crosses represents the uncertainty of the absolute position 
  of stars. 
}
\label{fig: EVN-MERLIN ToO 2010}
\end{figure}

Figure~\ref{fig: EVN-MERLIN ToO 2010} presents the pre-self-calibration map of 
the LHC emission of the 2010s' flaring event, obtained with the 
EVN-(\emph{e})MERLIN array, just past the OH maximum, in February 2010 at the 
optical phase +0.26. It shows the dominating component which was used for 
self-calibration so as to improve the signal to noise of the datacube.
Gaussian fitting of this component leads to an astrometric position of 
  RA$_{\rm J2000} = 02^h19^m 20.8091^s$, 
  DEC$_{\rm J2000} = -02^{\circ} 58\arcmin 41.935\arcsec$.
The purple and green crosses give the estimated positions of $o$~Ceti and its 
companion Mira~B, respectively. These positions are extrapolated from 
Matthews \& Karovska's (2006) VLA imaging, taking into account the 
proper motion (van~Leeuwen 2007).
The 2010s' flaring region is located $\sim$200$\pm$40~mas 
(i.e., $\le$~20$\pm$4~AU) east of $o$~Ceti. \\

From a 2-D Gaussian fitting of the ALMA band 3 \& 6 continuum observations 
taken between the 17$^{th}$ and 25$^{th}$ of October 2014 and between the 
29$^{th}$ of October and the 1$^{st}$ of November 2014 respectively, Wittkowski 
et al. (submitted) determined the position of Mira~A to be 
%%%
$02^h19^m20.795016^s$ ($\pm 0.00006\arcsec$) 
$-02^{\circ}58\arcmin43.03984\arcsec$ ($\pm 0.00007\arcsec$) and 
$02^h19^m20.795055^s$ ($\pm 0.00003\arcsec$)
$-02^{\circ}58\arcmin43.05078\arcsec$ ($\pm 0.00006\arcsec$) 

%%%
for the Band~3 and Band~6 epochs, respectively. Taking into account the 
proper motion of $o$~Ceti (van Leeuwen, 2007), these newer sets of 
position agree within  
20~mas to the one we inferred from Matthews \& Karovska's (2006) maps.
Adding quadratically the uncertainties means that the relative position 
of the MERLIN/EVN-(\emph{e})MERLIN maser components relative to the position of 
$o$~Ceti has an overall uncertainty of 40~mas. 
We interpolated the position of Mira~B relative to 
that of Mira~A adopting the measurement values compiled by 
Planesas, Alcolea \& Bachiller (2016). We adopted a conservative uncertainty 
of 40~mas, corresponding to the sum of the biggest individual-measurement 
uncertainty claimed by the authors and half the inferred positional shift 
in RA and DEC over the relevant period of the work presented here.

%%%%%%%%%%%%
%%%%%%%%%%%%
%%%%%%%%%%%%
%%%%%%%%%%%%

%%%%%%%%%%%%
%%%%%%%%%%%%
\begin{figure}
 \epsfig{file=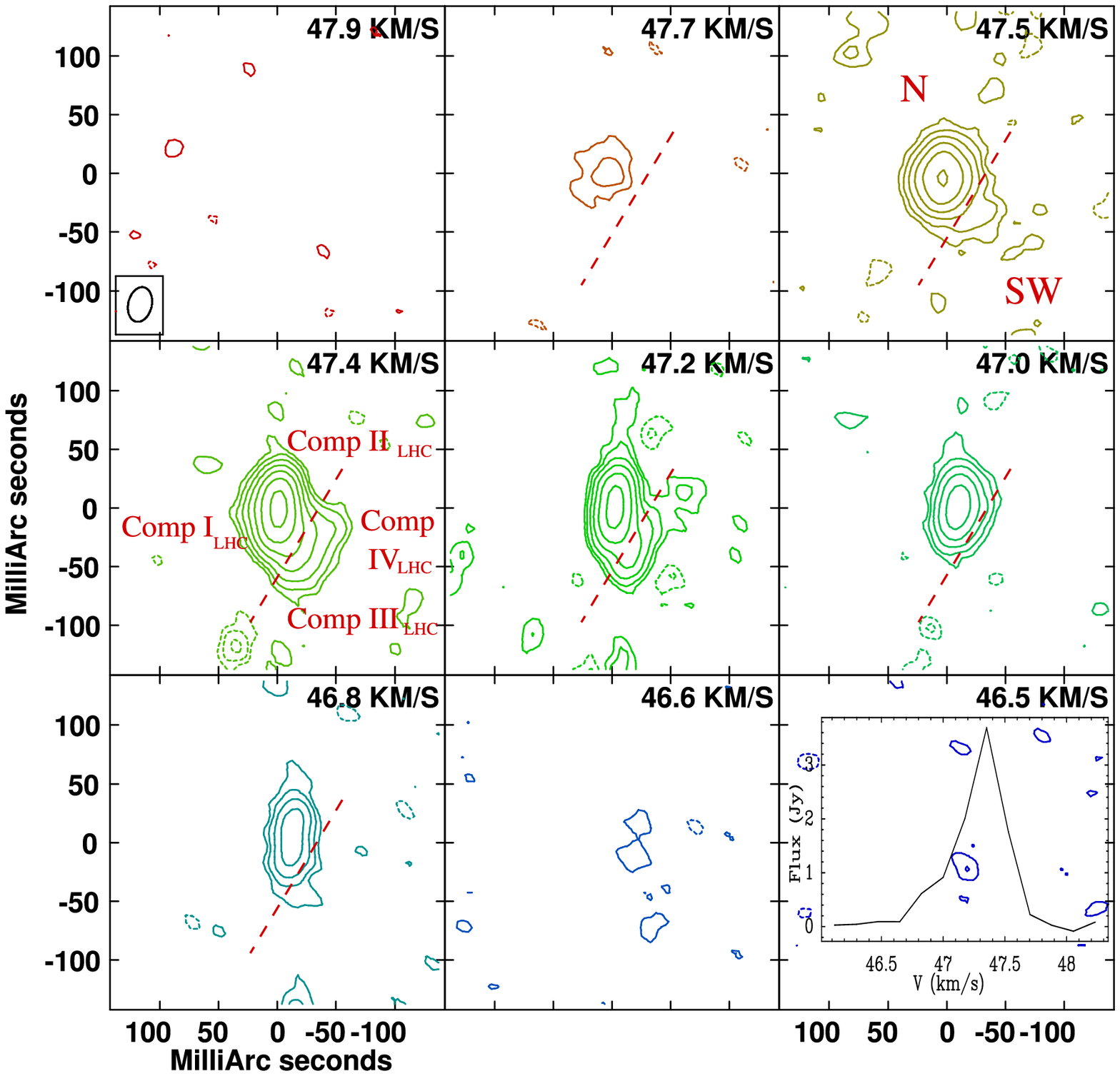,width=8.10cm,angle=-90}
\caption{1665-MHz LHC contour maps for all the channels where emission was 
detected in the EVN-(\emph{e})MERLIN observations. The restoring beam 
(0.03000~$\times$~0.02000~arcsec$^2$ with a PA=$-13$) is shown in the 
lower-left corner of the top-left channel map. The contours are at 
($-1$, 1, 2, 3, 5, 8, 10, 25, 35, 80) $\times$ 0.01828~Jy~beam$^{-1}$. 
The spectrum constructed from the datacube is superimposed over the 
lower-right channel map. The dashed line marks the border between the 
northern (N) and the south-west (SW) groups of components.}
\label{fig: 1665 LHC}
\end{figure}
%%%%%%%%%%%%
%%%%%%%%%%%%

%%%%%%%%%%%%
%%%%%%%%%%%%
\begin{figure}
 \epsfig{file=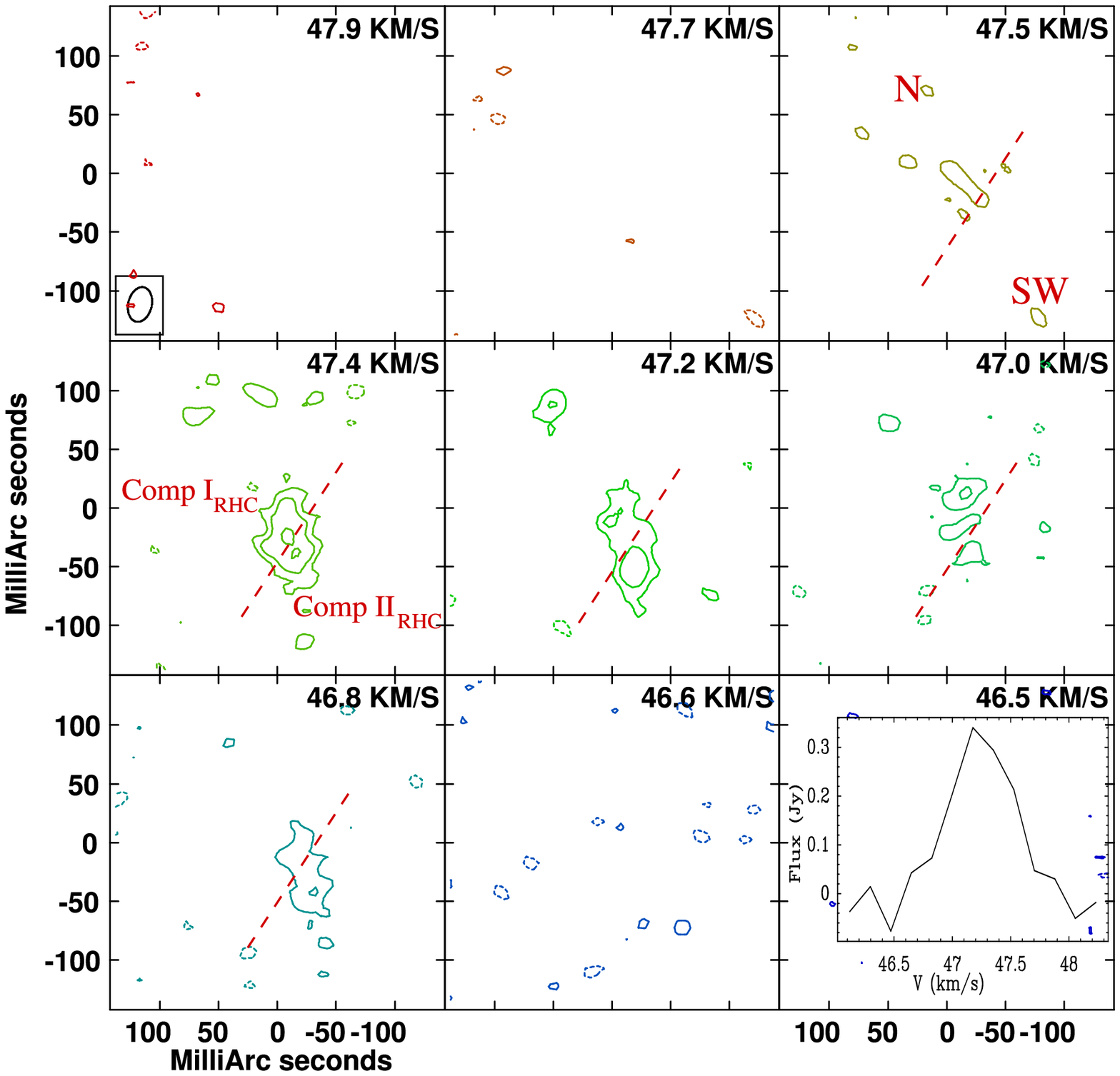,width=8.10cm,angle=-90}
\caption{Same as Fig.~\protect\ref{fig: 1665 LHC} for the 1665-MHz RHC. 
         The same restoring beam and contour levels have been used as 
         given in Fig.~\protect\ref{fig: 1665 LHC}.}
\label{fig: 1665 RHC}
\end{figure}
%%%%%%%%%%%%
%%%%%%%%%%%%

%%%%%%%%%%%%
%%%%%%%%%%%%
\begin{figure}
 \epsfig{file=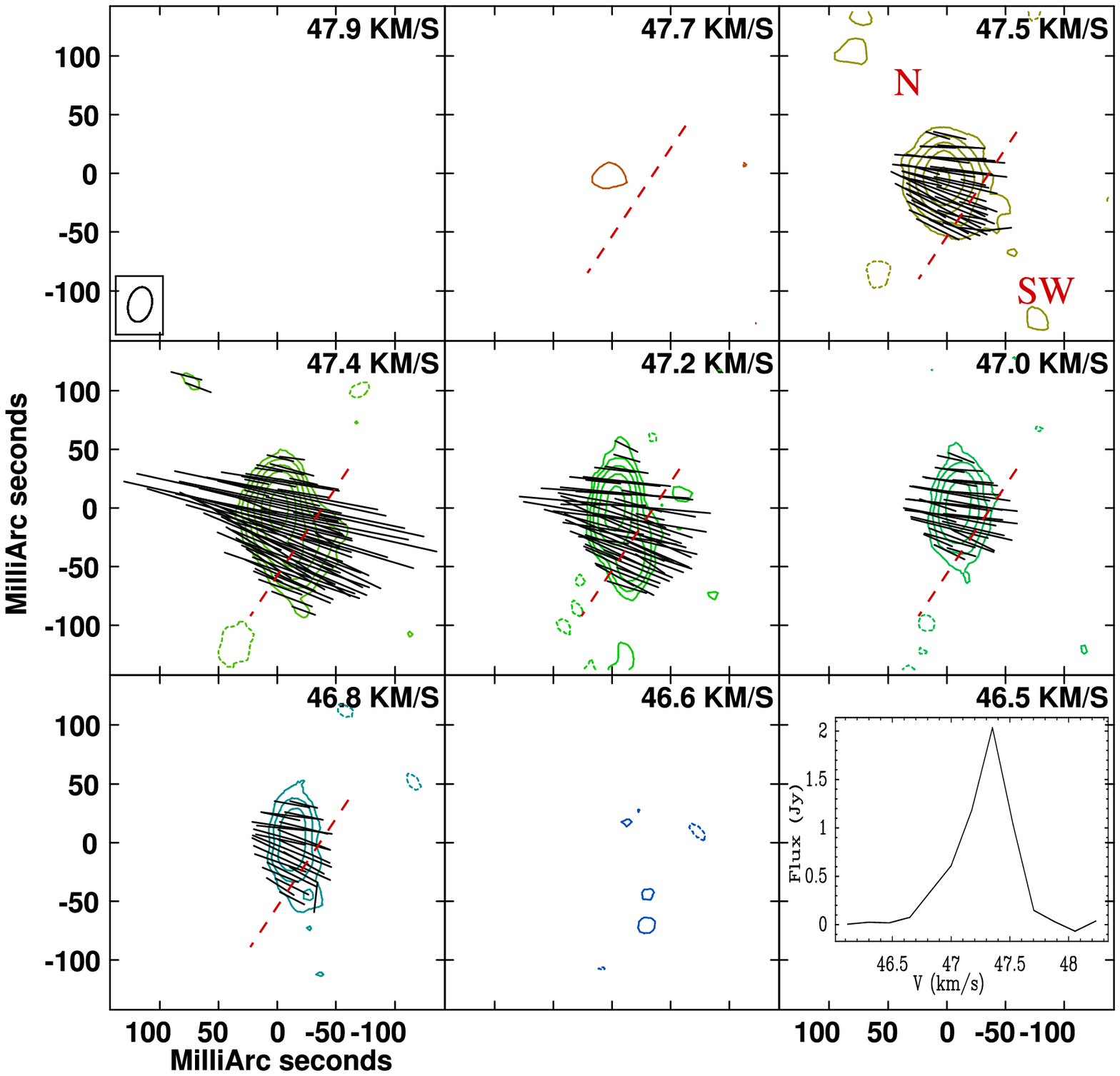,width=8.10cm,angle=-90}
\caption{EVN-(\emph{e})MERLIN 1665-MHz Stokes~{\it I} contour maps on which the 
         polarimetric information is overlaid. Note that this shows only the 
         relative linear PPAs as the observations of the PPA calibrator 
         3C286 were unfortunately unsuccessful. The length of a vector is 
         proportional to the percentage of linear polarisation. As in
         Figs~\protect\ref{fig: 1665 LHC}~\&~\protect\ref{fig: 1665 RHC}, 
         the spectrum constructed from the datacube is superimposed over the 
         lower-right channel map. The same restoring beam and contour levels 
         have been used as given in Fig.~\protect\ref{fig: 1665 LHC}.}
\label{fig: 1665 IQU}
\end{figure}
%%%%%%%%%%%%
%%%%%%%%%%%%

Figures~\ref{fig: 1665 LHC}~\&~\ref{fig: 1665 RHC} present the 
post-self-calibration contour maps of all the channels where emission was 
detected in the LHC and RHC polarisations, 
respectively. Also shown, superimposed on the lower-right channel map is the 
spectrum constructed from the respective datacubes. The comparison of the 
profile and peak intensity of the EVN-(\emph{e})MERLIN spectra with the 
average NRT single-dish spectra (cf. Fig.~\ref{fig: NRT Inte OH 1665}) 
indicates that at least 90\% of the signal has been recovered during the 
interferometric observations. \\

A close analysis reveals that in the LHC polarisation there are actually 
2 main groups of components: 2 strong components which are barely 16~mas apart 
from each other (labelled group ``N'') and 2 fainter components, $\sim$40~mas 
south-west of the 2 strong northern group of components 
(labelled group ``SW''). For visualisation purpose, a dash line marking the 
separation between the ``N'' and ``SW'' groups is displayed in the figures for 
the channels where emission was detected.
The 2 strong components in group ``N'' correspond to Comp~I$_{\rm \, LHC}$ \& 
Comp~II$_{\rm \, LHC}$ as labelled in 
Figs~\ref{fig: NRT Inte OH 1665}~\&~\ref{fig: NRT OH 1665 all component}.
In this group of components, there is a gradual shift from the western 
component (Comp~II$_{\rm \, LHC}$) to the eastern one (Comp~I$_{\rm \, LHC}$), 
while the velocity increases over the entire velocity range covered by the 
LHC polarised emission. 
The ``SW'' group of components (which spectral signature is not apparent in 
the averaged spectra shown in Fig.~\ref{fig: NRT Inte OH 1665} and that we 
shall label Comp~III$_{\rm \, LHC}$ \& Comp~IV$_{\rm \, LHC}$) span over the much 
smaller velocity range V=[$+47.1$,$+47.6$]~km~s$^{-1}$
(cf. Table~\ref{Table: Components and Zeeman pairs summary} 
presenting a summary of the velocity and location of the compoments).
Due to the faintness of the signal in RHC polarisation, the structure of the 
rather compact emission is less well-defined. Yet, both Comp~I$_{\rm \, RHC}$ 
and Comp~II$_{\rm \, RHC}$ (as labelled in
Figs~\ref{fig: NRT Inte OH 1665}~\&~\ref{fig: NRT OH 1665 all component})
can be identified in Fig.~\ref{fig: 1665 RHC}. 
Comp~I$_{\rm \, RHC}$, spanning the velocity range 
V=[$+47.0$,$+47.6$]~km~s$^{-1}$, is located in the western part of 
group ``N''. Comp~II$_{\rm \, RHC}$, spanning the velocity range 
V=[$+46.7$,$+47.4$]~km~s$^{-1}$ belongs to group ``SW''. 
Gaussian fitting was used to measure precisely the 
position of the various components to search for possible Zeeman pairs.
The criteria for a Zeeman pairing is a positional agreement to within 
the absolute positional uncertainty of 
($35 \times 10$)~mas$^2$ (cf. Section~\ref{sub: position accuracy}).
The Gaussian fitting reveals that in group ``N'', with a positional agreement 
of ($16 \times 8$)~mas$^2$, the western component is actually a Zeeman pair 
made of Comp~II$_{\rm \, LHC}$ (centred at $V_{\rm II _{LHC}} \sim +47.0$~km~s$^{-1}$) 
and its Zeeman counterpart Comp~I$_{\rm \, RHC}$ (centred at 
$V_{\rm I _{RHC}} \sim +47.4$~km~s$^{-1}$).
Note that the actual position of 
Comp~II$_{\rm \, RHC}$ is better visually guessed in the channel map  
$V = +47.0$~km~s$^{-1}$ as, in particular in the channel map 
$V = +47.4$~km~s$^{-1}$ it is hard to disentangle visually Comp~I$_{\rm \, RHC}$ and
Comp~II$_{\rm \, RHC}$.

In group ``SW'', Comp~III$_{\rm \, LHC}$ and Comp~II$_{\rm \, RHC}$ are also a 
Zeeman pair, but of much fainter intensity, preventing us to ascertain its 
velocity signature. \\
%- - - - - - - - - - - - - - - - - - - - - - - - - - - - - 
\begin{table}
\caption{\small Component characteristics summary}
\label{Table: Components and Zeeman pairs summary}
  \begin{tabular}{lcc}
\hline
Component            &  $^a$Peak Velocity {\em or}  & Zeeman Pairing  \\
Label                &  Velocity Range              &                 \\
                     &  (km~s$^{-1}$)                &                 \\
\hline
{\bf Group ``N''}    &                              &                 \\
Comp~I$_{\rm \, LHC}$   &  $V_{\rm Peak} \simeq +47.4$    &                \\
Comp~II$_{\rm \, LHC}$  & $V_{\rm Peak} \simeq +47.0$     & Z$_1$           \\
& \\
Comp~I$_{\rm \, RHC}$   & $V_{\rm Peak} \simeq +47.4$    & Z$_1$            \\
\hline
{\bf Group ``SW''}   &                             &                  \\
Comp~III$_{\rm \, LHC}$ &  [$+47.1$,$+47.6$]          & {\em $^{b}$Z$_2$} \\
Comp~IV$_{\rm \, LHC}$  &  [$+47.1$,$+47.4$]          &                  \\
 & \\
Comp~II$_{\rm \, RHC}$  &  [$+46.7$,$+47.4$]          & {\em $^{b}$Z$_2$} \\
\hline
 \end{tabular} \\
{\small 
{\bf a}: The peak velocity of the component is given when measurable with 
enough precision, else the velocity range of the component is given \\
{\bf b}: The faintness of the signal does not allow the measurement of the 
component peak velocities with enough precision preventing to ascertain 
the velocity signature of this Zeeman pair
}
\end{table}
%- - - - - - - - - - - - - - - - - - - - - - - - - - - - - 

Figure~\ref{fig: 1665 IQU} presents the Stokes~{\it I} contour 
maps on which the polarimetric information is overlaid. 
Note that as the observations of the PPA calibrator 3C286 
were unfortunately unsuccessful, the polarimetric vectors presented in the 
figure are not the absolute PPAs which are not retrievable. 
Only the difference in angles (i.e., the relative PPAs) are to be taken into 
consideration. Bearing this in mind, it is nonetheless clear that the 
distribution  of the vectors of polarisation attests to an underlying ordered 
but relatively complex magnetic field. The EVN-(\emph{e})MERLIN polarised 
maps show that the vectors of polarisation associated with group ``N'' and 
group ``SW'' have PPAs differing by $~\sim 20^{\circ}$ hence revealing the 
intricacy of the magnetic field lines probed by the flaring OH maser emission 
down to a resolution of a few tenths of mas.

%- - - - - - - - - - - - - - - - - - - - - - - - - - - - - - - - - - - - - - -
\subsubsection{The single-dish Medicina and Effelsberg H$_2$O monitoring}
\label{sec: H2O monitoring}
\noindent

\begin{figure}
 \epsfig{file=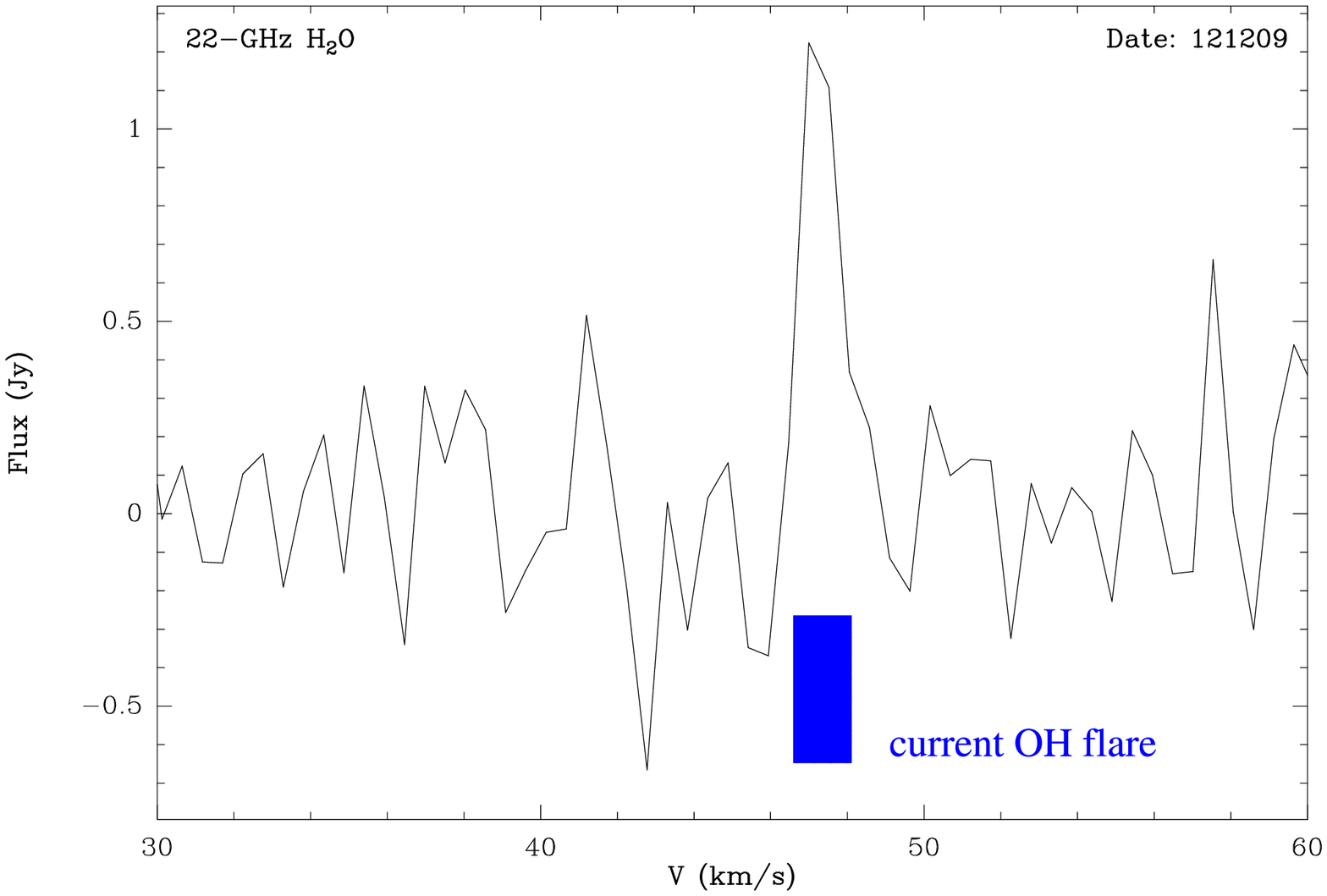,width=5.6cm,angle=-90}
\caption{Medicina spectrum of the H$_2$O maser emission obtained in December 
         2009 during the rise of the OH maser emission towards the maximum
         (integration time: 20 min, resolution: 0.53~km~s$^{-1}$; rms: 0.25~Jy).
         The width of the vertical bar gives the 2010s' OH emission velocity 
         spread.}
\label{fig: Medicina H2O spectrum}
\end{figure}

Figure~\ref{fig: Medicina H2O spectrum} presents a spectrum of the 22~GHz 
H$_2$O maser emission obtained in December 2009 with the Medicina antenna 
during the rise of the OH maser emission towards the maximum, on which the 
velocity range of the 2010s' OH flare is also displayed. It is clear that 
both maser species are emitting in a similar velocity range. What is more, 
they both peak at a similar velocity. \\

%%%%%%%%%%%%
%%%%%%%%%%%%
%%%%%%%%%%%%
%%%%%%%%%%%%

%%%%%%%%%%%%
%%%%%%%%%%%%
\begin{figure}
\epsfig{file=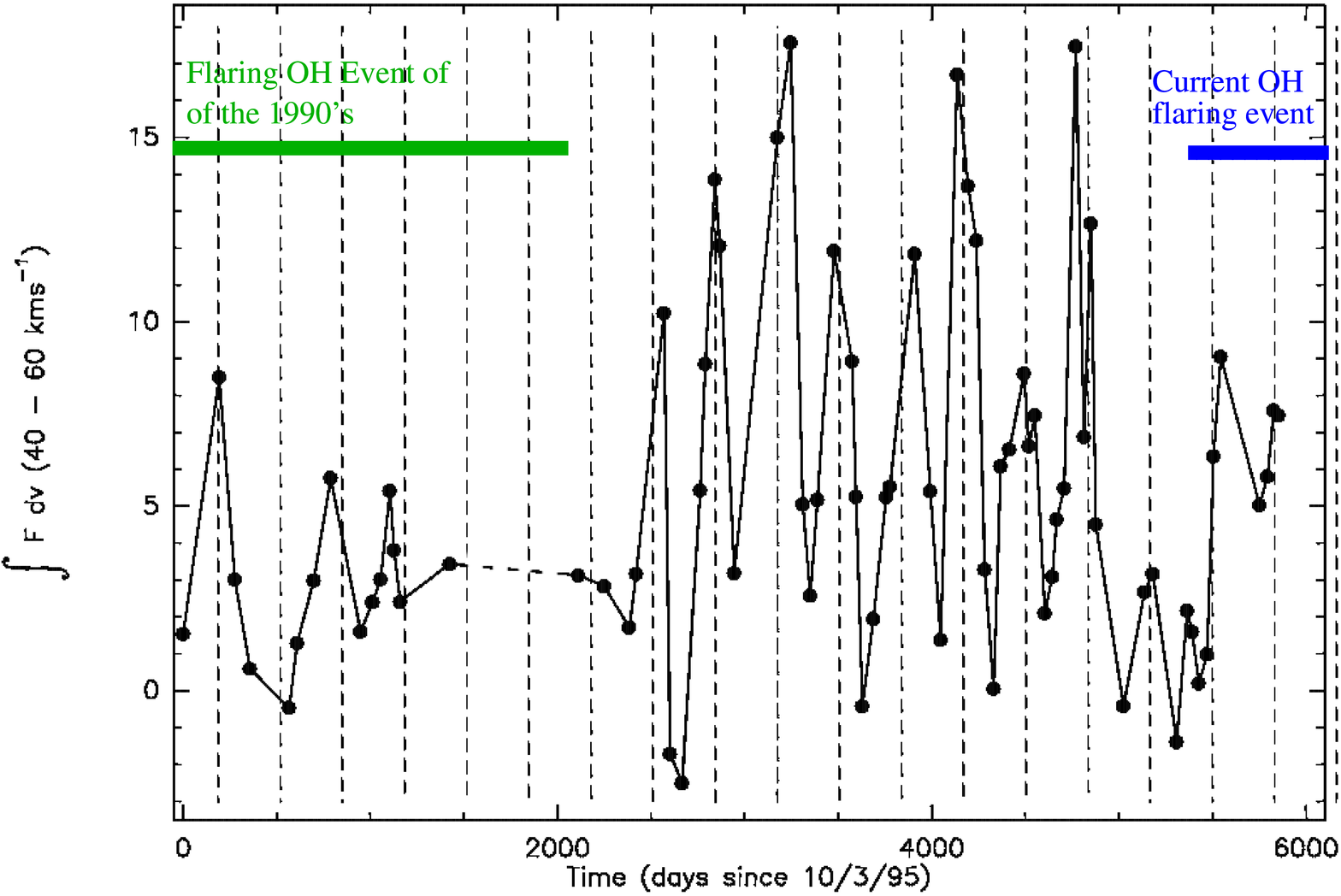,width=5.60cm,angle=-90}
\caption{Variability curve of the 22-GHz water maser emission recorded from 
March 1995 until March 2011 at Medicina (68 spectra) and Effelsberg (6 spectra).
Note: no observations were performed between the 01/03/1999 and 20/12/2000 
represented by the nearly horizontal dashed line. 
The vertical dashed lines mark the 332-day optical period
arbitrarily shifted to roughly coincide with peaks 
in the H$_2$O emission), which the water maser seems to follow quite well. 
The OH flaring events recorded during the time interval presented here are 
also displayed.
}
\label{fig: H2O long term variability}
\end{figure}
%%%%%%%%%%%%
%%%%%%%%%%%%
Figure~\ref{fig: H2O long term variability} shows the variability curve of the 
22-GHz water maser emission from March 1995 until March 2011 based 
on  68 and 6 spectra taken with the Medicina and Effelsberg telescopes 
respectively. The vertical dashed lines mark the 332-day optical period, 
which the water maser seems to follow quite well. 
Note though that the lines have been arbitrarily shifted to roughly 
coincide with peaks in 
the H$_2$O emission (and do not mark the actual optical minima or maxima).  
They serve as a guide to show that the optical period characterises 
reasonably well also the H$_2$O periodicity. 
Further detailed analysis of the H$_2$O variability 
characteristics, which is beyond the scope of this article, will be presented 
in Brand et al. (2017).
Also displayed in the figure are the two OH maser flaring events of the 
1990s and the 2010s.
Interestingly, the comparison of the long-term variability of the 22-GHz 
H$_2$O maser emission with that of the OH maser activity towards $o$~Ceti, 
shows that OH flaring events seem to appear when the 22-GHz H$_2$O is 
relatively fainter. 

%----------------------------------------------------------------------
\section{Discussion}
\label{sec: discussion}

\subsection{Location of the flaring regions}
\label{sub:flaring Geometrical structure}

The 1995 MERLIN observations and the EVN-(\emph{e})MERLIN 
observations of the current 2010s' flare were obtained 14~years 
2~months apart. Therefore, a proper motion of 
$\delta{\rm PM}_{\rm RA} \simeq +0.132\arcsec$ and 
$\delta{\rm PM}_{\rm Dec}\simeq -3.36\arcsec$ is expected.
The difference measured between the position of the maser component observed 
in the 2010s' event and the strongest one in the 1995 map is 
$\delta_{\rm RA} \simeq +0.621\arcsec$  
$\delta_{\rm Dec} \simeq -3.24\arcsec$, while it is 
$\delta_{\rm RA} \simeq +1.35\arcsec$  
$\delta_{\rm Dec} \simeq -3.84\arcsec$ with respect to the faintest  
component of the 1995 map.
The 2010s' maser position is consequently offset from the positions of the 
stronger and fainter maser components observed in 1995 by $0.5\arcsec$ and 
$1.3\arcsec$, respectively. This is significantly greater than the expected 
distance travelled over 14~years by the same parcel of material in the CSE 
due to expansion, hence implying that the 1990s' flaring emission and the 
2010s' one originate from 2 distinct regions 
(cf. also the discusion regarding the flare velocity properties in 
section~\ref{sub: velocity information}). \\

\begin{figure}
\epsfig{file=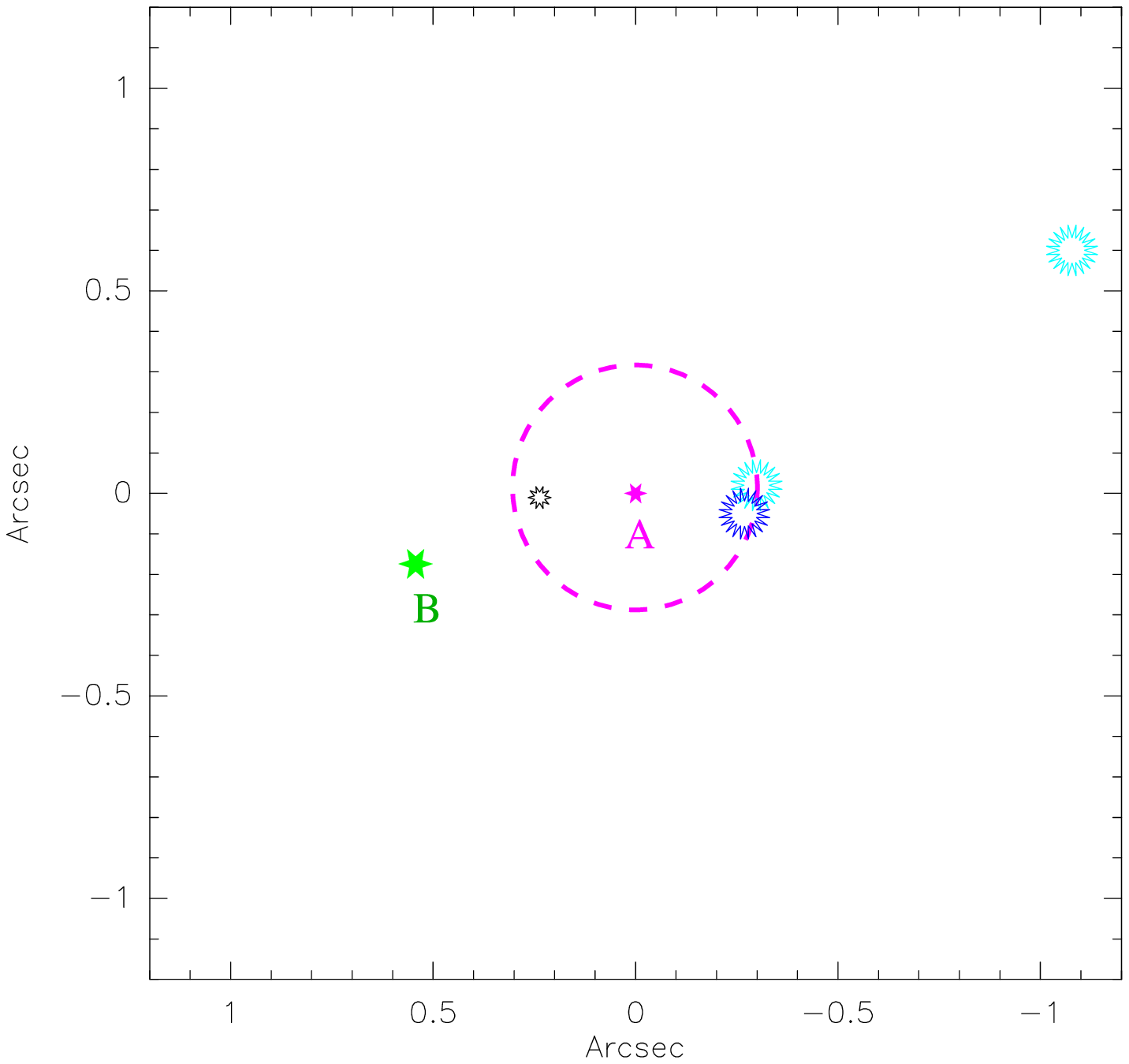,width=7.90cm,angle=-90}
\caption{Relative positions of the OH flaring region measured by 
the EVN-(\emph{e})MERLIN in February 2010 (black), the MERLIN 
masers (light blue: 1995 / dark blue: 1998) and the stars of the Mira~AB 
system (purple star: Mira~A /green star: Mira~B), taking Mira~A as the 
reference position, and correcting all the positions for proper motion. 
Note that the size of the symbols represent the astrometric uncertainty of 
the given positions for the stars while for the masers, it represents their 
convolved sizes. The dashed-circle centred on $o$~Ceti and giving 
the distance to the centre of the strongest 1995 maser component has a 
radius of $0.3\arcsec$ ($\sim$30~AU).
}
\label{fig: Masers and stars relative positioning}
\end{figure}

Figure~\ref{fig: Masers and stars relative positioning} presents the 
relative positions of the OH flaring region measured by the 
EVN-(\emph{e})MERLIN in 
February 2010 along with the MERLIN masers detected in 1995 and 1998 and 
the stars of the Mira~AB system, taking Mira~A as the reference position, 
and correcting all the positions for proper motion 
(cf. Table~\ref{Table: maser and MiraB relative position} for the calculated 
offsets). 
While the 2010s' flaring event is clearly affecting the region located 
between Mira~A and Mira~B, the 1990s' event affected only the part of the 
shell in the opposite side to Mira~B. \\

%- - - - - - - - - - - - - - - - - - - - - - - - - - - - - 
\begin{table}
\caption{\small Offsets of the maser components and Mira~B relative to 
                $o$~Ceti}
\label{Table: maser and MiraB relative position}
  \begin{tabular}{lccllll}
\hline
                                  & X$_{\rm offset}$    & Y$_{\rm offset}$   \\
                                  &   ($\arcsec$)    &  ($\arcsec$)    \\
\hline
  Mira B                          &     +0.543       &   $-0.174^{*}$   \\
  MERLIN 95 strongest maser comp. &     $-0.299$     &       +0.017    \\
  MERLIN 95 faintest maser  comp. &     $-1.078$     &       +0.596    \\
  MERLIN 98 maser comp.           &     $-0.269$     &       $-0.052$  \\
  EVN-(e)MERLIN maser spot        &     +0.240       &       $-0.014$  \\

\hline
 \end{tabular} \\
Reference position: extrapolated position of $o$~Ceti the 04$^{th}$ of May 
1998 inferred to be  RA$_{\rm J2000}=02^h19^m20.7860^s$, 
                     DEC$_{\rm J2000}=-02^{\circ}58\arcmin39.130\arcsec$ \\
*: Relative position to Mira~A, interpolated from the series of 
   measurements compiled by Planesas, Alcolea \& Bachiller (2016)
\end{table}
%- - - - - - - - - - - - - - - - - - - - - - - - - - - - - 

%%%%%%%%%%%%
%%%%%%%%%%%%
%%%%%%%%%%%%
%%%%%%%%%%%%

In order to better constrain the exact position of the flaring regions in 
relation to Mira~B, a more constrained determination of the orbital movement  
is needed. In particular, the ALMA observations of late October-early November 
2014 lead to a most modern measurement of the separation of the Mira~A and B of 
0.472~arsec (corresponding to 43.4~AU at a distance of 92~pc) and a  hint 
towards a decrease in separation (Vlemmings et al. 2015). 
Still, with this uncertainty in mind (i.e., $\sim$100~mas of uncertainty for 
the absolute positioning of Mira~B), it is striking, that 
the strongest maser emission detected in the 3 epochs originate from a 
projected distance of less than 0.4$\pm 0.04\arcsec$ 
(that is less than 40$\pm$4~AU, taking into consideration the convolved size 
of the maser components while the dashed-circle in 
Fig.~\ref{fig: Masers and stars relative positioning} gives the distance to 
the centre of the strongest 1995 maser component) from Mira~A, with a 
hint of a potentially ``deeper'' OH flaring region on the side of the 
companion. 
Only a very faint maser component is found further out, $\sim 1.1\arcsec$ 
(i.e.,  $\sim$110~AU) away. 
This strongly suggests that the projected distances measured over these 
3 independent epochs are a good estimation of the actual radius at which the 
recurrent flaring events appear, which is unusually close to the central 
star. \\

There is a relation between the OH radius and the mass-loss rate
(and hence the shell thickness, Huggins \& Glassgold 1982) as well as  
between the expansion velocity observed in the OH main lines of Miras 
and their period (Sivagnanam et al. 1989). So, one expects the OH (main-line) 
standard shell size to be roughly related to the period of the central Mira, 
which is indeed in agreement with the typical sizes found by mapping 
(Chapman, Cohen \& Saikia 1991; Chapman et al. 1994).
In particular, Chapman, Cohen \& Saikia (1991) measured the radius of the 
loci of the strongest OH emission around U~Ori, a thin-shell Mira 
with a similar period duration (368 days, Kukarkin et al. 1970) located at 
$\sim$306~$\pm$61~pc (Mondal \& Chandrasekhar (2005) using 
Whitelock \& Feast's (2000) Period-Luminosity relation),
to be $0\farcs21-0\farcs24$. From these mapping results, 
we can infer the typical radius at which the bulk of the standard OH 
main-line  maser emission for a Mira having a period of about one year is 
expected to arise from, to be about R$_{peak}\sim$100~AU (with fainter emission 
extending at smaller and greater radius around the R$_{peak}$) in agreement 
with the location of the fainter remote OH maser component detected here. \\

The faint more distant maser component observed in 1995, is then most likely 
coming from the standard OH shell. 
The fact that only one such maser component was observed and only at one epoch, 
suggests that further out in the CSE, in particular at the location of 
the standard OH envelope, the conditions for maser emission do not seem to 
be optimal. 
Furthermore, we note that this remote faint maser component is located on the 
opposite side of the shell with respect to the companion Mira~B. The presence 
of the close companion is most likely to strongly perturb the standard CSE 
part of the shell to the side closer to it, inhibiting maser emission in these 
regions. \\

%- - - - - - - - - - - - - - - - - - - - - - - - - - - - - - - - - - - - - 
\subsection{Flare propagation and velocity drift}
\label{sub: velocity information}

The propagation of the flaring region during the 1990s' 
event was happening at a speed of V$\sim$25~km~s$^{-1}$ with a hint of a 
clockwise propagation. With such a velocity and direction of 
propagation, the 1990s' flaring region would be expected to be roughly 
south of $o$~Ceti after the 12.5 years separating the 1990s' and 2010s' 
flaring events. \\

Due to its binary nature, $o$~Ceti is expected to undergo an orbital motion 
around the barycentre of the system. Baize (1980) proposed a first estimation 
of the orbital elements of the system which have been recalculated more 
recently by Prieur et al. (2002), in the light of new speckle observations 
of $o$~Ceti. Nevertheless, and as pointed out by the latter authors,  
since even the sum of the masses is unknown and the trajectory is still 
poorly sampled, the orbital elements of the binary system can not be 
tightly constrained. 
The observed velocity drift of $\sim$[0.27~--~0.35]~km~s$^{-1}$ 
in 888~days of the peak of the maser emission 
(that is $\sim$[0.11~--~0.14]~km~s$^{-1}$ per year), could be the signature 
of the orbital motion of Mira~A around the barycentre of the system.
Due to the play of intensity of the spectral components during time, 
there is an indication that the velocity drift of the emission peak 
is most likely higher than the intrinsic drift of each individual 
spectral component. A more refined value of the intrinsic velocity drift would 
require the complete analysis of the long term variability characteristics 
of the individual spectral component, which is beyond the scope of this paper. 
Such an analysis will be presented in a subsequent paper.
Bearing in mind this latter warning and that a wide range of parameters is 
possible in terms of orbital parameters, adopting the most recent orbital 
period determination of Prieur et al. (2002; P$\sim$500~yr), a total mass 
of the system of 3.0~M$_\odot$, a typical Mira mass of 1.0~M$_\odot$ for $o$~Ceti,
a deprojected separation of $\sim$85.6~AU between the 2 stars would lead to 
an intrinsic acceleration of $\sim$0.05~km~s$^{-1}$~yr$^{-1}$ around 
the barycentre of the system for $o$~Ceti. 
Note that the total mass of 3.0~M$_\odot$ adopted here would not necessarily 
preclude Mira~B from being a white dwarf, as the 2.0~M$_\odot$ represents the 
overall counterpart mass of the binary system, corresponding to Mira~B and the 
surrounding material of the binary system. 

%- - - - - - - - - - - - - - - - - - - - - - - - - - - - - - - - - - - - - 
\subsection{Front or back part of the CSE? \\ 
            {\em A stellar velocity determination issue}}
\label{sub: stellar velocity determination issue}

Several AGB stars show a double-component profile in CO, composed of a 
relatively strong narrow component superimposed on a fainter and 
broader component, with both components centred on the same velocity
(Knapp et al., 1998; Winters et al 2003). 
Such a profile is well constrained by a two-component parabolic fit and is 
interpreted by the aforementioned authors as a ``double-wind'' signature 
(i.e., the signature of 2 steady winds).
$o$~Ceti is one of the stars that show 2 components in both 
its CO(3$-$2) and CO(2$-$1) profiles. But, since both lines 
clearly lack the red-shifted emission of the putative broad component, 
their profiles do not seem to be well constrained by such a composite 
narrow-broad component model. 
Assuming the validity of this model for $o$~Ceti -- that is the existence of 
a broad component centred at the same central velocity as the narrow 
component -- would imply that the red part of the broad component is missing, 
advocating for a highly asymmetrical ``external'' shell. Another 
possible interpretation which could account for the asymmetric line profile 
observed in the case of $o$~Ceti, has been put forward by Josselin et al. 
(2000). 
They mapped the CO(2$-$1) with the IRAM interferometer and also obtained 
optical spectroscopic observations of the KI lines in an attempt to detect the 
faint outer parts of the CSE. They propose the existence of a spherical shell 
disrupted  by a bipolar outflow. The idea of a possibly bipolar CSE 
was previously proposed by Knapp \& Morris (1985). \\

\begin{figure}
  \epsfig{file=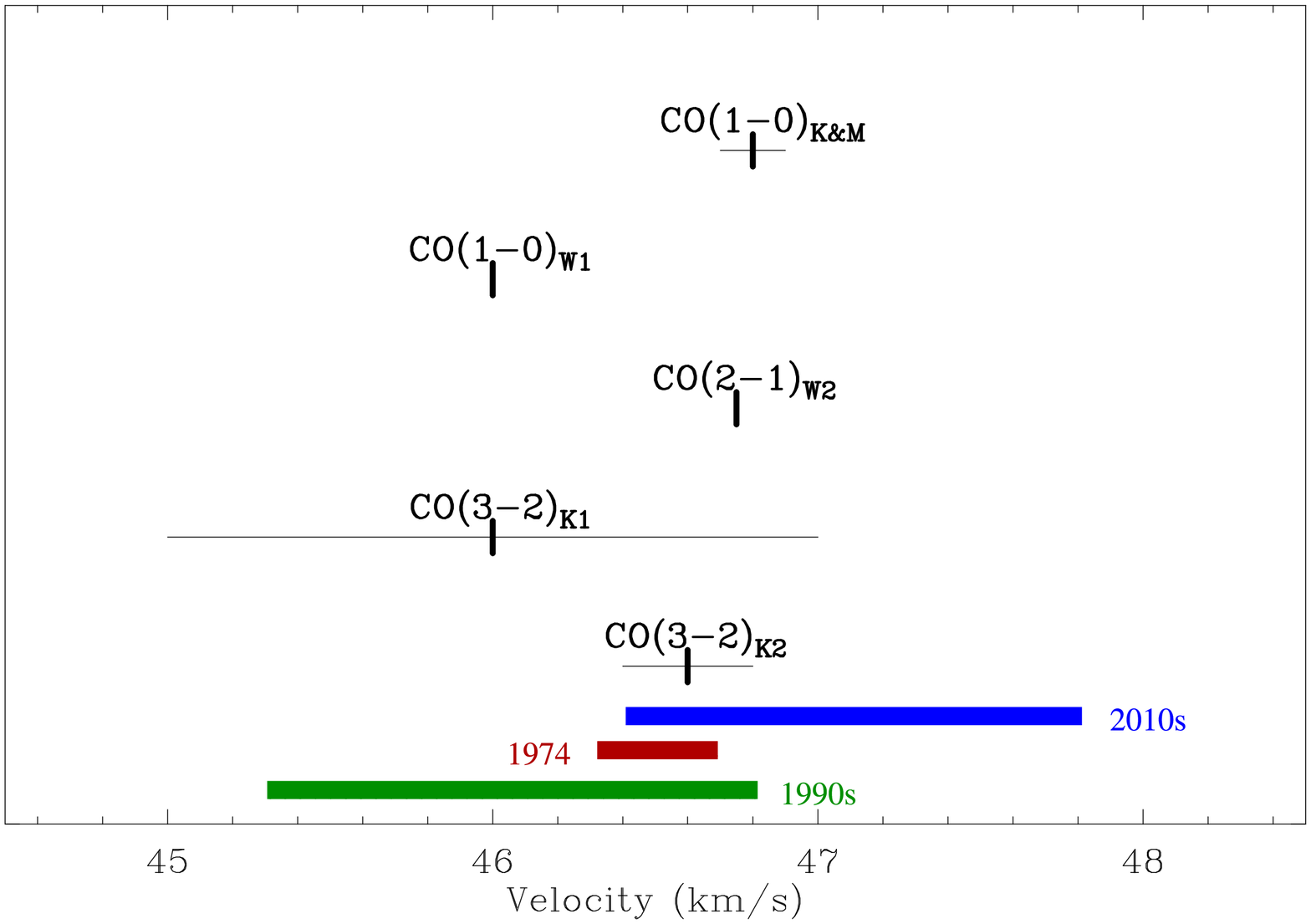,width=5.60cm,angle=-90}
\caption{Diagram showing the various estimates of the stellar velocity from CO
measurements and the OH maser velocity spread observed in 1974 (Dickinson, 
Kollberg \& Yngvesson 1975), the 1990s and the 2010s. Uncertainties for the 
CO velocities are represented by horizontal bars. ``{\bf K\&M}'' stands for 
Knapp \& Morris (1985), ``{\bf K}'' stands for Knapp et al. (1998) and  
``{\bf W}'' stands for Winters et al. (2003). Note that the 2 values,  
``{\bf K1}'' and ``{\bf K2}'' from Knapp et al. (1998) come from the 
same transition due to the two-component parabolic fit used by the authors 
to account for the spectral profile; though
a similar two-component parabolic fit was also used by Winters et al. (2003)
they used a fixed stellar velocity within a given transition. 
Note also that Winters et al. (2003) do not provide any uncertainties, 
but a step of 0.25~km~s$^{-1}$ has been used for the fits (Le~Bertre, 
private communication). 
}
\label{fig: Mira CO and OH Velocities}
\end{figure}

The non-symmetrical two-component nature of the CO profiles of $o$~Ceti 
reveals the complexity of its CSE with clear signs of asymmetry 
present even in the outermost part, but it also makes the 
determination of the stellar velocity and the final expansion velocity 
more problematic.
Knapp \& Morris (1985) estimate the stellar velocity from their CO(1$-$0)
spectra to be V$=+46.8 \pm 0.1$~km~s$^{-1}$. Knapp et al. (1998), using a 
two-component parabolic fit (but not forcing the broad and narrow components
to have the same central velocity) 
estimated the central velocity of the 2 components of the CO(3$-$2) emission to 
be V$=+46.6 \pm 0.2$~km~s$^{-1}$ and V$=+46.0 \pm 1.0$~km~s$^{-1}$ for the narrow 
and broad components respectively. 
Winters et al. (2003), also using a 
composite parabolic fit (but imposing a single central velocity for 
both the narrow and broad components of the profile), estimated the stellar 
velocity to be V$=+46.0$~km~s$^{-1}$  and  V$=+46.75$~km~s$^{-1}$ from the 
CO(1$-$0) and the CO(2$-$1) profiles respectively 
(note that no uncertainties are given for these estimations, but a step of 
0.25~km~s$^{-1}$ has been used for the fits; Le~Bertre, private communication). 
Considering all these estimations and their attached uncertainties, 
this leads to a rather loosely constrained 
stellar velocity, within the range [$+45.0$,$+47.0$]~km~s$^{-1}$ 
(cf. Fig.~\ref{fig: Mira CO and OH Velocities}). 

Note that independent inference of the systemic velocity from 
the inner part of the CSE (few R$_*$) via SiO 
emission, adopted to be 45.7~$\pm$~0.7~km~s$^{-1}$ 
(Cotton et al. 2006) 
or more recently of $46.7$~km~s$^{-1}$ by Wong et al. (2016), 
based on complex asymmetrical profiles, produce a similar range 
for the stellar velocity of $o$~Ceti. \\

With such a velocity range, encompassing that of the 1990s' 
(and the mid-1970s') flaring one, it is currently not 
possible to decide whether the flare in the 1990s and that of the 
mid-1970s (within the range V$\simeq$~[$+46.4$,$+46.9$]~km~s$^{-1}$ and 
centred at $+46.5$~km~s$^{-1}$, respectively) emanates from in front or behind 
the star. Nonetheless, with a velocity range for the current 2010s' 
flare (V$\simeq$[$+46.5$,$47.9$]~km~s$^{-1}$) it is likely that it is 
located in the back part of the shell.

%%%%%%%%%%%%
%%%%%%%%%%%%
%%%%%%%%%%%%
%%%%%%%%%%%%

%- - - - - - - - - - - - - - - - - - - - - - - - - - - - - - - - - - - - - 
\subsection{Comparison with previously recorded flaring events}
\label{sub: previously flaring events}

So far, 6 other flaring events in thin-shell Miras have been recorded.
The first record of such an event was made towards U~Ori at 1612~MHz 
(Jewell, Webber \& Snyder, 1981). Five other events were then recorded, at 
1612~MHz towards U~Her and R~LMi, and in the main lines towards X~Oph, 
R~Leo, and R~Cnc (Etoka \& Le~Squeren, 1996, 1997). A study based on all  
6 objects was presented in Etoka \& Le~Squeren (1997).
All 6 objects are found in a delimited portion of the [$60-25$]~vs~[$25-12$] 
IRAS colour-colour diagram where the bulk of non-OH oxygen-rich Miras are found.
Even though the duration of the flares varies from a few months to several 
years, they are all characterised by a very short rise time. The flaring 
feature is always characterised by 
$|$V$_{\rm star} - $V$_{\rm flare}| < $~V$_{\rm OH \; exp}$ 
(where V$_{\rm flare}$ is the peak velocity of the flaring feature, 
V$_{\rm OH \; exp}$ is the standard OH expansion velocity and V$_{\rm star}$ the 
stellar velocity) which is clearly related to the [$25-12$] colour. The 
flaring emission shows substantial polarisation. \\

The OH flaring features in $o$~Ceti also show a strong polarisation. 
Due to the unusual nature of its OH emission, only observed as 
flaring events, 
a direct estimation of $o$~Ceti's standard OH expansion velocity 
is not possible. The Sivagnanam et al. (1989) Velocity-Period relation 
implies a V$_{\rm OH \,exp} \sim$3--5~km~s$^{-1}$ 
for the period of $o$~Ceti (cf. Section~\ref{sub: 1990 MERLIN maps}).
Chapman et al. (1991) estimated a V$_{\rm OH  \,exp} \sim$~7~km~s$^{-1}$ for U~Ori. 
As noted previously, U~Ori and $o$~Ceti have about the same period 
and hence should have a comparable expansion velocity. 
Knowing that the terminal expansion velocity of the CSE of $o$~Ceti, measured 
in the CO transitions, is only V$_{\rm CO  \,exp} \sim~6 \pm 0.2$~km~s$^{-1}$ 
(Knapp \& Morris 1985) and that a small acceleration is still present at the 
location of the OH shell of Miras (e.g., Chapman et al. 1994), it is likely 
that for $o$~Ceti V$_{\rm OH \; exp} \simeq$[4~--~5]~km~s$^{-1}$. This means that 
the flaring emission of $o$~Ceti is 
also characterised by $|$V$_{\rm star} - $V$_{\rm flare}| < $~V$_{\rm OH \; exp}$ since 
$|$V$_{\rm star} - $V$_{\rm flare}| <$~3~km~s$^{-1}$, regardless of the actual 
stellar velocity (cf. Fig~\ref{fig: Mira CO and OH Velocities}). \\

\begin{figure}
  \epsfig{file=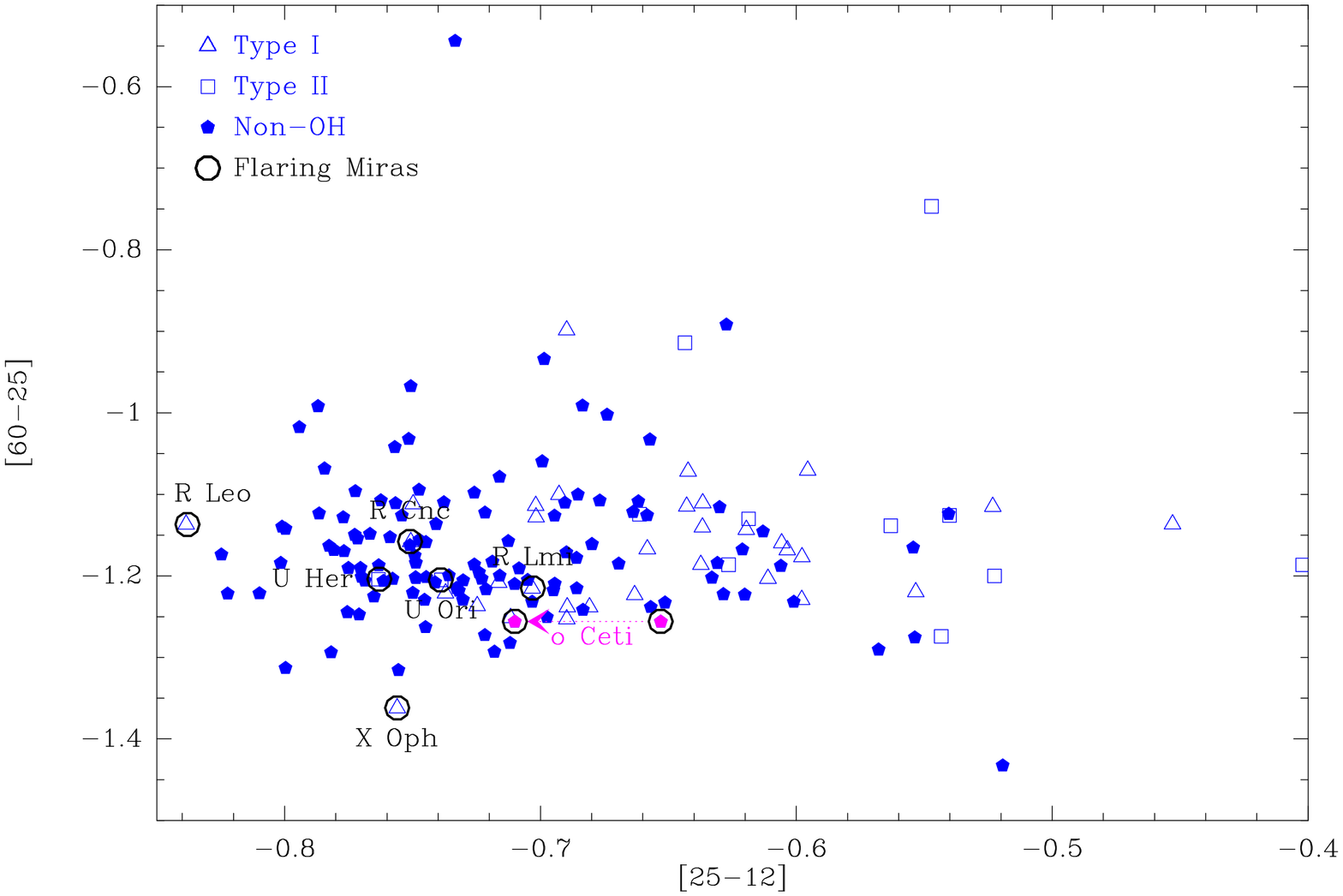,width=5.70cm,angle=-90}
\caption{Adapted from Fig.~10 of Etoka \& Le~Squeren (1997) showing the
[$60-25$]~vs~[$25-12$] IRAS colour-colour diagram of the nearby Miras 
(distance $<$~1~kpc) with the location of all the flaring Miras indicated 
including $o$~Ceti. ``Type I'' stands for Miras which are emitting 
predominantly in the 1665/67~MHz main lines; ``Type~II' stands for Miras which 
show their strongest emission in the 1612-MHz satellite maser line and 
``non-OH'' stands for Miras which have not been detected in any of the 
ground-state OH maser transitions. 
Note that 2 positions are given for $o$~Ceti in this diagram. The ``redder'' 
one corresponds to the uncorrected value as calculated from the fluxes given 
in the IRAS Catalogue (i.e., corresponding to the overall binary system), 
while the ``bluer'' value is corrected for the mid-infrared excess identified 
by Ireland et al. (2007) and interpreted by the latter authors as 
being due to the accretion disk heated by Mira~A (cf. text). 
The arrow shows the direction of the correction.
}
\label{fig: IRAS colour-colour flaring objects}
\end{figure}

Figure~\ref{fig: IRAS colour-colour flaring objects} is the adapted
Fig.~10 of Etoka \& Le~Squeren (1997) showing the [$60-25$]~vs~[$25-12$] IRAS
colour-colour diagram of the nearby Miras (distance $<$~1~kpc) with the 
location of all the flaring Miras indicated including $o$~Ceti. 
It shows the locations of ``non-OH'' Miras, that is the Miras which have not 
been detected in any of the ground-state OH maser transitions, the ``Type I'' 
Miras which are emitting predominantly in the 1665/67~MHz main lines and 
the ``Type~II'' Miras, which have thicker CSEs and show their strongest 
emission in the 1612-MHz satellite line.

%%%%%%%%%%%%
%%%%%%%%%%%%
%%%%%%%%%%%%
%%%%%%%%%%%%

It has to be noted though that the fluxes given in the IRAS catalogue 
correspond to the overall Mira~AB system and hence comprise also an excess of 
contribution due to the presence of the 2 stars in the beam, 
making $o$~Ceti look slightly ``redder'' than it actually is.
Ireland et al. (2007) performed some mid-infrared observations of the Mira~AB 
system and interpreted the mid-infrared excess they observed as coming from an 
optically-thick accretion disk heated by Mira A. Using the model fits of their 
work, constraining the SED of this disk (associated with Mira~B), between 0.35 
and 18.3~$\mu$m and doing a simple extrapolation to 25~$\mu$m, one can 
estimate its contamination into the IRAS measurements at 12 and 25~$\mu$m.
This leads to a corrected [25-12] IRAS colour ranging between 
-0.690 to -0.710 (as opposed to an uncorrected value of -0.653). 
Adopting such a correction, the [$25-12$] IRAS colour for $o$~Ceti is indeed 
in agreement with the identified ``flaring Mira area'' of 
Etoka \& Le~Squeren's (1997) former work. \\

\begin{figure}
  \epsfig{file=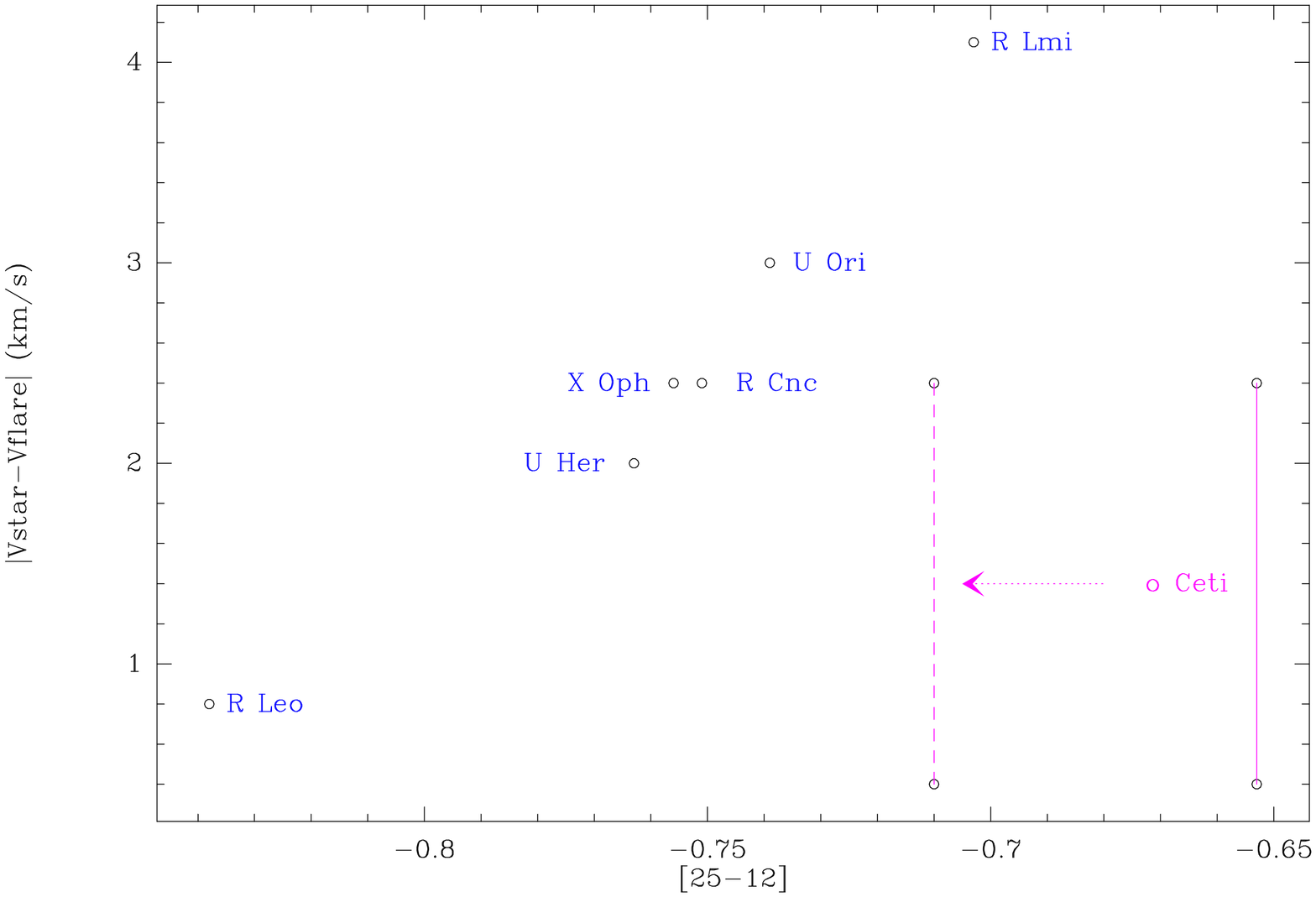,width=5.70cm,angle=-90}
\caption{Adapted Fig.~11 of Etoka \& Le~Squeren (1997) showing the
relation found between $|$V$_{\rm star} -$V$_{\rm flare}|$ and the [$25-12$] 
IRAS colour of the flaring Miras and the location of $o$~Ceti in this 
diagram. Note that the actual location of $o$~Ceti in this diagram is 
rather poorly constrained due to, on the one hand, the contamination of its 
[$25-12$] colour by the presence of its companion Mira~B and, in the other 
hand, its poorly constrained stellar velocity, illustrated here by the 2 
vertical lines taking into accounts these 2 effects.
}
\label{fig: V_flare vs [25-12]}
\end{figure}

Figure~\ref{fig: V_flare vs [25-12]} presents the adapted Fig.~11 of 
Etoka \& Le~Squeren (1997) showing the relation found between 
$|$V$_{\rm star} - $V$_{\rm flare}|$ and the [$25-12$] IRAS colour of the 
flaring Miras, with the location of $o$~Ceti in this diagram. 
The actual location of $o$~Ceti in this diagram is rather 
poorly constrained due to, on the one hand, the contamination of its 
[$25-12$] colour by its companion Mira~B 
and, on the other hand, its poorly constrained stellar velocity. 
This is illustrated in the 
diagram by the 2 vertical lines taking into account these 2 effects.
Considering the decontaminated [$25-12$] colour brings $o$~Ceti to a 
better agreement with the [$25-12$] vs $|$V$_{\rm star} - $V$_{\rm flare}|$ 
relation observed for the rest of the flaring Miras. Yet, the location of 
$o$~Ceti in this diagram indicates that its $|$V$_{\rm star} - $V$_{\rm flare}|$ 
value is smaller than anticipated from the relation, even for the 
most optimistic decontaminated [$25-12$] colour. This hints at an even deeper 
location of the flaring region. A possible interpretation of this is that it 
reflects the influence of the companion ``carving'' even deeper the 
flaring zone in the side of the shell where it resides, by supplying an 
extra amount of anisotroptic UV radiation.

%%%%%%%%%%%%
%%%%%%%%%%%%
%%%%%%%%%%%%
%%%%%%%%%%%%

%- - - - - - - - - - - - - - - - - - - - - - - - - - - - - - - - - - - - - 
\subsection{OH luminosities of flaring Miras}
\label{sub: Luminosity limit for maser emission}

Nguyen-Q-Rieu et al. (1979) performed a statistical study towards 48 Miras 
to determine their OH intrinsic luminosities. 
They find that the Type-I OH intrinsic luminosity 
ranges from $\sim$5$\times 10^{15}$ to $\sim$5$\times 10^{17}$~Watt~Hz$^{-1}$, 
while the Type-II OH intrinsic luminosity ranges from $\sim 10^{17}$ to 
$\sim$5$\times 10^{18}$~Watt~Hz$^{-1}$.
A more recent statistical study performed over the $>$2000 Galactic 
stellar sources known to be OH emitters (i.e., including Miras, SRs and 
OH/IR stars) from the catalogue of Engels \& Bunzel (2015) was made by
Etoka et al. (2015). Discarding the few low- and high-luminosity outliers 
in the distribution, the upper limits found in this new study agree with 
those of Nguyen-Q-Rieu et al. (1979) but the lower limits extend much further 
for both main-line and 1612-MHz satellite-line emitters by more than 2 
orders of magnitude. 
Taking the mean intensity of the OH emission observed towards each flaring 
Mira, leads to an OH intrinsic luminosity ranging from 
$\sim$2$\times 10^{12}$ to $\sim$3.5$\times 10^{13}$~Watt~Hz$^{-1}$, with that of 
$o$~Ceti being the lowest. This range of luminosities corresponds to the 
lower part of the distribution in Etoka et al. (2015) and might 
be taken as typical for the flaring stellar maser population.
This Mira group represents then the lower range of the Mira population both in 
terms of luminosity and mass-loss rate. From these 2 characteristics and their 
location in the [$60-25$]~vs~[$25-12$] IRAS colour-colour diagram, i.e., 
in the midst of the non-OH Miras and the edge  of the Type-I Miras, one 
plausible interpretation is that this group represent a transition between 
the 2 populations, pinpointing the lower limits in 
terms of physical properties needed for the OH maser to be present in the CSE.

%- - - - - - - - - - - - - - - - - - - - - - - - - - - - - - - - - - - - - 
\subsection{Polarisation and underlying magnetic field structure}
\label{sub: standard CSE model implication}

Etoka \& Le~Squeren's (1996, 1997) spectral records of the 
LHC and RHC spectra of the previous flaring events occuring in 
the 1665-MHz transition clearly show hints of Zeeman splitting signatures 
but a confirmation and accurate measurement of the associated magnetic field 
strength was not possible since no maps of these events were available. \\

Though the absolute polarisation angle associated with the maser components
could not be retrieved, preventing us from making a detailed analysis of the 
underlying magnetic field structure, it is clear from the distribution  of the 
linear vectors of polarisation presented in Fig.~\ref{fig: 1665 IQU} that the 
masers in $o$~Ceti trace an ordered but relatively complex magnetic field 
structure. The Zeeman pair detected in group ``N'' 
(cf. Table~\ref{Table: Components and Zeeman pairs summary}), 
at a distance from the star $\sim$200$\pm$40~mas (i.e., $\le 20 \pm 4$~AU), 
leads to a magnetic field of $B \sim +0.68$~mG, using the Zeeman splitting 
coefficient given by Davies (1974). \\

The high polarisation along with the erratic variability behaviour observed 
in the OH main-line flaring Miras is reminiscent of what is observed towards 
semiregular variable stars (Etoka et al. 2001; Szymczak et al. 2001).
This type of stars is characterised by light curves less regular than that 
of Miras and the variation in their optical amplitudes is less than 
2.5$^{\rm m}$ (Kholopov et al. 1985). They have infrared properties quite close 
to those of Miras emitting predominantly in the 1665/67~MHz main lines 
(i.e., the ``Type-I'' maser emitters). Analysis of their long-term OH maser 
emission variability and polarisation properties suggests that these 
characteristics are due to transient instabilities in their hot and thin CSEs 
(Etoka et al. 2001) where turbulence effects in the circumstellar magnetic 
field as well as magnetic field structural change are thought to occur 
(Szymczak et al. 2001).

%- - - - - - - - - - - - - - - - - - - - - - - - - - - - - - - - - - - - - 
\subsection{Implication of the flare location with respect to the standard 
            (OH) CSE model}
\label{sub: standard CSE model implication}

From a statistical analysis of all the Miras previously recorded to have 
exhibited OH flaring events, Etoka \& Le~Squeren (1997, cf. also 
Section~\ref{sub: previously flaring events}) concluded that the 
flaring emission is likely to originate from a region closer to the 
star than the distance at which OH maser emission in the standard model comes 
from.
As mentioned in Section~\ref{sec: H2O monitoring}, the velocity peak and spread 
of the OH flaring emission and the 22-GHz H$_2$O are the same.
This additional clue strongly supports the suggestion that the OH maser 
flaring emission observed here indeed originates from a region closer to 
the star than the standard OH maser CSE distances, in agreement with the 
findings of Etoka \& Le~Squeren (1997).  
Furthermore, as also mentioned Section~\ref{sec: H2O monitoring}, the long term 
monitoring of the 22-GHz H$_2$O maser emission shows that OH flaring events 
seem to appear when the 22-GHz H$_2$O is relatively fainter.
A tentative explanation for this behaviour is that it is the signature of 
an enhanced OH production by photodissociation of H$_2$O which translated 
itself by a decrease of the water maser emission. \\

OH flaring events close to the central star only seem to occur in 
thin-shell Miras, indicating that the physical conditions for 
the masers to occur in these more internal zones are not fulfilled
for thick-shell stars, possibly due to a non-favourable 
combination of temperature and/or density and/or pumping conditions. \\

Goldreich \& Scoville (1976) and Huggins \& Glassgold (1982) studied the 
physical properties of the CSE, demonstrating the importance of the ambient 
interstellar UV radiation in the production of OH molecules by 
photodissociation of H$_2$O. Huggins \& Glassgold (1982) show the variation of 
the peak of the OH density profile formed by photoproduction with respect to 
the mass-loss rate and the importance of H$_2$O shielding in this process.
While Goldreich \& Scoville (1976) show that such a process delivers an 
important source of OH in the outer part of the CSE, their results also show 
that a high abundance of OH is expected in the CSE close to the star 
(cf. their Fig.~4, though admittedly their models are more adequate for 
OH/IR objects due to the high mass-loss rates adopted:
$3 \times 10^{-5}$~M$_\odot$~yr$^{-1}$). \\ 

Moreover, Cimerman \& Scoville (1980) studied the possible importance of 
direct stellar radiation at 2.8~$\mu$m for the pumping scheme of OH maser lines 
(particularly effective for the main lines) in late-type stars. Using 
Goldreich \& Scoville (1976) CSE models to test this scheme, 
Cimerman \& Scoville suggest the existence of two zones of 
high OH emissivity with such a pumping mechanism where IR pumping plays a 
more significant role for the zone nearer to the star. Also,  
Collison \& Nedoluha (1993) stress that the NIR pumping scheme proposed 
by Cimerman \& Scoville (1980) can operate at significantly lower column 
densities of OH than the FIR pumping scheme. \\

While some faint 1667~MHz emission was observed during the 1990s' event, 
emission at that frequency was not detected during the time interval covering 
the current 2010s' flaring event presented here. The fact that 1665~MHz is 
excited while 1667~MHz is only sporadically observed is in favour of a denser 
or/and warmer environment. 
Indeed, from a statistical study of main-line emission towards Miras, 
Sivagnanam et al. (1989) show that the expansion velocities at 1665 and 
1667~MHz are such that $\Delta$V$_{1667} > [1.1 - 1.3] \Delta$V$_{1665}$,  
indicating that acceleration is still present at the location where these maser 
transition occur and that 1667-MHz maser emission extends further out in the 
OH shell than the 1665-MHz maser emission (confirmed by mapping by e.g.
Chapman et al. 1991). 
Furthermore,  models (Elitzur 1978, Bujarrabal et al. 1980) show that higher 
gas temperature and density are more favourable to 1665-MHz maser 
emission implying that 1665-MHz maser emission is expected to be found down to 
a slightly more internal radius than 1667-MHz emission. \\

%%%
On the whole, the different positions of the OH masers between 1995 and 2010 
(cf. Fig.~\ref{fig: Masers and stars relative positioning}) fall within a 
circle of radius $0.3\arcsec$, comparable with the SiO emission presented in 
Wong et al. (2016). Even the weak OH feature at $1\arcsec$ observed in 
1995, seems to have a counterpart in the extended SiO emission.
Also, we incidentally note that the excited H$_2$O transition at 232.67~GHz,
mapped by ALMA, with an excitation energy 
$E_{up}/k \sim 3462$~K (Wong et al. 2016), hence tracing the warm H$_2$O 
layer, is confined within a much smaller radius of $\leq 0.1\arcsec$ from 
O~Ceti than the OH flaring region.

%- - - - - - - - - - - - - - - - - - - - - - - - - - - - - - - - - - - - - 
\subsection{Role of binarity in the flaring events towards $o$~Ceti}
\label{sub: role of binarity}

Danchi et al. (1994) measured the inner radii of the dust shell of a sample of 
13 late-type stars, including $o$~Ceti, at 11.15~$\mu$m. They found two 
classes of stars. The group which $o$~Ceti belongs to, is the one for which 
the stars have their dust shells very close to the photosphere 
($\sim$3 stellar radii for $o$~Ceti).
They observed  $o$~Ceti at a wide range of optical phases and with 3 
different baselines, one of which (the 13~m baseline for which the position 
angle of PA=113$^\circ$) is virtually aligned with the Mira~AB axis 
(PA=$112^\circ \pm 1^\circ$, Karovska, Nisenson \& Beletic 1993) and shows 
visibilities that hint at a smaller radius than for the other baselines. 
The authors also notice a similar signature of asymmetry in the TiO, which 
peaks closer to the star on the side illuminated by the companion Mira~B. \\

Similarly, Lopez et al. (1997) performed visibility observations at 
~11~$\mu$m towards $o$~Ceti during a $\sim$7-year period, in the 
Mira-AB axis direction,  
which stress the non-spherical and evolving clumpy structure of the dust 
shell around $o$~Ceti.
They even suggested that the formation of a new clump of dust could be
linked to the OH flaring events observed in the 1990s by increasing 
temporarily the far-infrared emission which is known to operate as a common 
pump of the OH ground-state main lines (i.e., 1665 \& 1667~MHz). 
Finally, very recent mapping of the CO emission around the Mira~AB system 
with ALMA (Ramstedt et al. 2014) revealed the presence of a bubble, which 
the authors tentatively suggest is created by the wind of Mira~B and blown 
into the expanding envelope of Mira~A. \\

The general flaring characteristics observed for $o$~Ceti are similar to those 
observed for flares in isolated Miras in terms of IRAS colour-colour 
properties, velocity range in relation to the stellar velocity, spectral 
characteristics and polarisation behaviour as described by 
Etoka \& Le~Squeren (1997), which are quite distinct 
from what is observed towards standard Miras (Etoka \& Le~Squeren 2000).
These common characteristics suggest that the flaring locations in terms of 
radius in the CSE is similar for all these events and all indicate that these 
regions of transient OH maser activity are distinct from the standard OH maser 
zone.
Yet, in the case of $o$~Ceti there are also hints (e.g., the nature of 
the companion itself, asymmetrical emission in e.g. the CO and TiO, and 
the relatively small $|$V$_{\rm star} - $V$_{\rm flare}|$  value) that 
the presence of the companion is playing a role in the flares, 
most probably in producing ``episodic'' H$_2$O photodissociation in 
preferential regions.  \\

%----------------------------------------------------------------------
\section{Summary and Conclusions}
\label{sec: conclusion}

We have presented the analysis of the onset of the new 2010s' flaring event 
currently undergone by $o$~Ceti and compared its characteristics with those
of the 1990s' flaring event, based on a series of single-dish and 
interferome\-tric complementary observations both in OH and H$_2$O, obtained 
with the NRT, Medicina and Effelsberg telescopes and the MERLIN and
EVN-(\emph{e})MERLIN arrays. 
We also compared the overall characteristics of $o$~Ceti's recurring flaring 
events with those which have been observed towards other thin-shell Miras and 
summarized by Etoka \& Le Squeren (1997). 

The NRT monitoring shows that the spectral profile of the 1665-MHz flaring 
emission of $o$~Ceti is comprised of a set of 2 main persistent components 
both in the LHC and RHC polarisations. For the 2010s' flaring event period 
presented here, the well-sampled monitoring shows that the maser emission 
follows the optical light curve variation with a delay of $\sim$70~days, 
corresponding to a phase delay of +0.2 as is typically observed towards 
variable Miras. 
While the spectral profile and the velocity spread observed in the 1990s' and 
2010s' event are simliar (i.e., consisting of 2 main spectral components 
with an overall velocity spread of 1.5~km~s$^{-1}$), the central velocity 
differs by $\sim$1~km~s$^{-1}$ between the 2 events. 
A velocity drift of $\sim$0.27--0.35~km~s$^{-1}$ in $\sim$900~days of the 
1665-MHz emission peak was observed during the 1990s which could be the 
signature of the movement of $o$~Ceti with respect to the barycentre of the 
binary system.

The multi-epoch mapping during the 1990s' event showed that the flaring region 
is not static but actually slowly propagates within the affected zone, 
at a speed of $\sim$25~km~s$^{-1}$.

Though both OH main lines are generally observed in standard Miras, only very 
weak 1667-MHz non-polarised emission was observed intermittently during 
the 1990s' event. 

During the 2010s' flaring event, the contemporary mo\-ni\-toring of 
the 22~GHz H$_2$O emission and the flaring OH maser emission revealed that 
emission of both species have a similar velocity range and peak at a 
similar velocity. The monitorings also revealed that OH flaring events seem 
to appear when the 22-GHz H$_2$O is relatively faint. This is interpreted as 
being due to a higher OH production due to an enhanced photodissociation of 
H$_2$O.

The 2010 high angular resolution EVN-(e)MERLIN maps revealed the presence
of 2 main groups of components. The polarised information of these data 
allowed us to infer that the flaring region is associated with a 
$B \sim +0.68$~mG relatively complex magnetic field. 

While the current 2010s' flaring event is located about $\sim$240$\pm$40~mas 
(i.e., $\le 24 \pm 4$~AU) east of $o$~Ceti between the 2 stars of the 
binary system, in the 1990s it was located at a similar distance from $o$~Ceti 
but on the other side of the star. 
Taking the 2 epochs into consideration, this leads to a radius for the flaring 
zone of less than $\sim$400$\pm$40~mas (i.e., $\le 40 \pm 4$~AU), with a 
hint of a potentially ``deeper'' OH flaring region in the side of the 
companion.  

The general flaring characteristics in terms of infrared 
properties, velocity of the flaring components (compared to the OH standard 
expansion velocity expected for these stars) and polarimetric behaviour, are 
similar to those recorded towards 6 other thin-shell Miras. 
The OH intrinsic luminosity deduced from the overall sample of flaring 
Miras ranges from
$\sim$2$\times 10^{12}$ to $\sim$3.5$\times 10^{13}$~Watt~Hz$^{-1}$,
with that of $o$~Ceti being the lowest. This range of luminosities might be 
taken as typical for the flaring stellar maser population.
$o$~Ceti's first multi-epoch interferome\-tric mapping of such events 
confirms the unusually short radius at which such events occur and hence 
the suggestion brought by Etoka \& Le~Squeren (1997) that OH flaring zones 
in thin-shell Miras are located closer to the star than the standard-OH 
maser-emission zone. 
A possible explanation of such maser zones could reside in the pumping scheme 
at play, with the predominance of NIR pumping for eruptive zones of thin-shell 
Miras, which works at lower column densities of OH than the FIR pumping 
known to be the main pumping scheme for the standard OH maser zone.

%----------------------------------------------------------------------
\section*{Acknowledgements}
{\footnotesize
The {Nan\c cay} Radio Observatory is the Unit\'e Scientifique de {Nan\c cay}
of the Observatoire de Paris, associated with the CNRS. The {Nan\c cay} 
Observatory acknowledges the financial support of the R\'egion 
Centre in France. The European VLBI Network is a joint facility of 
European, Chinese, South African and other radio astronomy institutes 
funded by their national research councils. 
MERLIN is a UK national facility operated by the 
University of Manchester on behalf of STFC.
The Medicina 32-m radiotelescope is operated by 
INAF-Istituto di Radioastronomia in Bologna.
We also acknowledge, with thanks, the use of data from the 
AAVSO (American Association of Variable Star Observers). 
We thank the referee, V. Bujarrabal, for his constructive comments 
and suggestions.
}
%----------------------------------------------------------------------

%---------------------------------------------------------------

\end{document}